\documentclass{elsarticle}

\usepackage{mathrsfs}
\usepackage{lineno,hyperref}
\usepackage{tikz}
\usepackage[export]{adjustbox}
\usepackage{amsmath}
\usepackage{geometry,tikz}
\usetikzlibrary{arrows,matrix,positioning,fit}
\usepackage{amssymb}
\usepackage{physics}
\usepackage{float}
\usepackage{subcaption}
\usepackage{array}
\usepackage{multirow}
\usepackage{multicol}
\usepackage{array}
\newcolumntype{P}[1]{>{\centering\arraybackslash}p{#1}}
\newcolumntype{M}[1]{>{\centering\arraybackslash}m{#1}}

\usepackage{caption}
\usepackage{relsize}
\usepackage[figuresleft]{rotating}
\usepackage{ulem}
\modulolinenumbers[5]

\journal{Ocean Engineering}

\newcommand{\fig}[1]{Fig.~\ref{#1}}
\newcommand{\eq}[1]{Eq.~(\ref{#1})}

\DeclareCaptionFormat{cont}{#1 #2#3\par}

\newcommand{\vx}{u} 
 
\newcommand{\vz}{w} 
\newcommand{\sx}{\eta_{,x}} 
 
\newcommand{\eL}{\eta^{(1)}}
\newcommand{\deL}{\dot{\eta}^{(1)}}
\newcommand{\vzL}{w^{(1)}} 
\newcommand{\sxL}{{\eta_{,x}}^{(1)}}
\newcommand{\vxL}{u^{(1)}} 
\newcommand{\de}{\dot{\eta}} 
\newcommand{\xvg}{{x_{1}}}
\newcommand{\yvg}{{x_{2}}}
\newcommand{\zvg}{{x_{3}}}  
\newcommand{\xv}{\chi}
\newcommand{\yv}{\dot{\chi}}
\newcommand{\zv}{\zeta}  
\newcommand{\yvi}{\tau}
\newcommand{\rh}{\rho}
\newcommand{\rhp}{\dot{\rho}}
\newcommand{\rhL}{\rho^{(1)}}
\newcommand{\rhpL}{\dot{\rho}^{(1)}}
\newcommand{\nuz}{\mu^{_\uparrow}}
\newcommand{\lc}{\ell}
\newcommand{\su}{{\sigma_{\eta}}}
\newcommand{\sd}{{\sigma_{\de}}}
\newcommand{\st}{{\sigma_{\xi}}}
\newcommand{\dep}{h}
\newcommand{\depa}{\tilde{h}_p}
\newcommand{\dnl}{W_{\rm nl}}

\begin{document}

\begin{frontmatter}

\title{Level-crossing distributions of kinematic variables in multidirectional second-order ocean waves}


\author[enstaAddress]{Romain Hasco\"{e}t}
\cortext[mycorrespondingauthor]{Corresponding author}
\ead{romain.hascoet@ensta-bretagne.fr}

\address[enstaAddress]{ENSTA Bretagne, CNRS UMR 6027, IRDL, 2 rue Fran\c{c}ois Verny, 29806 Brest Cedex 9, France}

\begin{abstract}

The conditional value of a stationary random process, given the level-upcrossing of another dependent stationary random process, is considered. Assuming that both processes are weakly non-Gaussian, an analytical approximation for the related conditional distribution is derived. It is based on a trivariate Edgeworth expansion truncated to non-Gaussian terms of lowest order, to which Rice's formula is then applied. As an application, the effect of level-upcrossing conditioning in second-order ocean waves is investigated. Upcrossing events are monitored for the sea surface elevation. The conditional distributions of different kinematic variables, given upcrossing, are considered for different sea-state configurations. Predictions from the analytical model are compared with numerical data obtained from Monte Carlo experiments. It is found that the analytical approximation provides conditional mean and variance in good agreement with numerical data, although moderate discrepancies appear for the sea states with the most severe wave steepnesses. Regarding the conditional skewness, given upcrossing, results are mixed, with significant discrepancies between the analytical approximation and numerical estimates, in a number of cases. An Edgeworth-type approximation is also provided for the upcrossing frequency and compared with Monte Carlo estimates; this analytical estimate is found to be accurate over a wide range of crossing levels.

\end{abstract}


\end{frontmatter}


\section{Introduction}

Stochastic level-crossing in the marine context finds various applications, 
such as slamming induced by wave impacts on a marine structure (e.g. \cite{ochi_1964a,ochi_1971,helmers_2012,hascoet_2020}) 
or the study of the random motions of the sea surface (e.g. \cite{longuet_1957,baxevani_2003}).
Level-crossing events may be monitored for the sea free-surface itself,
or for some response to wave excitations (e.g. the hull-girder response of a ship; see \cite{jensen_2001}).
When crossings are monitored for the sea surface, 
level crossing may be considered at a point fixed in space, along a line of sight at a given time,
or at a moving material point (e.g. a material point with seakeeping motions \cite{helmers_2012, hascoet_2020} 
or forward motion, \cite{lindgren_1999, aberg_2008,hascoet_2021}).\footnote{
More generally, level crossing may be also considered for a random field in a multidimensional space 
(for instance on contour lines for a two-dimensional random field; see \cite{baxevani_2003}).
}
Problems of practical interest may relate to the frequency of free-surface crossings,  
and/or the conditional distribution of wave-related variables, given free-surface crossing.
These problems can be addressed by the use of the level-crossing theory of random processes,
based on the pioneering work of Rice \cite{rice_1944,rice_1945} and subsequent developments 
(see for example \cite{lindgren_2012}, chapter 8, and references therein).

In most studies dealing with stochastic level-crossing in ocean waves, a linear wave model is assumed,
which allows the randomness of the free surface and wave-related variables to be modeled through Gaussian processes.
Then it offers a favorable framework, where level-crossing problems are mostly analytically tractable.
As soon as nonlinearities are introduced, the stochastic processes are no longer Gaussian,
and analytical developments become much more challenging.
Among the studies dealing with the statistics of nonlinear ocean waves,
most of them limit their investigations to the effect of second-order Stokes' corrections.

Based on the pioneering work of Kac and Siegert \cite{kac_1947,kac_1947b},
several studies investigated the probabilistic properties of a second-order response process through the 
eigendecomposition of the related Volterra kernel (see e.g. \cite{naess_1985, naess_1986, naess_1990, langley_1987}).
However, the generalization of this approach to multiple random processes 
-- as required for the present topic of crossing sampling --
is impeded by the fact that two interdependent random processes 
will have distinct Volterra kernels, 
which, \textit{a priori}, do not admit a common set of orthogonal eigenfunctions.
Series expansions of the probability distribution of the 
considered random processes,
is another approach which has been introduced by Longuet-Higgins to investigate the probabilistic effects 
of second-order nonlinearities in the context of ocean engineering (\cite{longuet_1963, longuet_1964}).
These series are named in honor of F. Y. Edgeworth \cite{edgeworth_1906}.
One advantage of Edgeworth series, compared to the eigendecomposition of the Volterra kernel, 
is that it can be readily generalized to multivariate distributions.
Besides, their applicability is not limited to second-order nonlinearities,
although accounting for higher-order nonlinearities remains very challenging in the context of ocean engineering.

The present study makes use of Edgeworth expansions to model 
the effect of second-order nonlinearities 
on wave statistics conditioned to free-surface upcrossing.
An upcrossing event is defined as the free-surface upcrossing a given level 
(i.e. a given altitude) at a position fixed in the reference frame of the mean flow
(i.e. the reference frame where the mean fluid velocity field is zero).
Based on a trivariate Edgeworth expansion, an analytical approximation is derived
for the conditional distribution of a second-order wave variable, given free-surface upcrossing.
The Edgeworth-type approximation of the second-order upcrossing frequency --
whose expression was already derived by Longuet-Higgins \cite{longuet_1964} -- is also considered.
For illustrative purpose, three different kinematic variables are considered: 
(i) the vertical component of the fluid velocity, 
(ii) the horizontal component of the fluid velocity
and (iii) the free-surface slope.
The analytical approximations obtained for their respective conditional distributions, given upcrossing, 
are compared with numerical results from Monte Carlo simulations of second-order irregular seas.
The analytical approximation of the second-order upcrossing frequency is also compared with numerical estimates.
These comparisons are used to delimit the applicative scope of the analytical approximations.
Section \ref{sec_notations} introduces the notations and conventions used throughout the paper.
Section \ref{sec_edgeworth} sets the theoretical framework of Edgeworth series 
and their application to second-order ocean waves.
Section \ref{sect_rice_ew} applies Rice's formula to Edgeworth series truncated to the leading order,
in order to obtain approximate formulae for the upcrossing frequency 
and the conditional distribution of a wave variable given free-surface upcrossing.
Section \ref{sec_illust_examples} illustrates the analytical model through a series of examples
and assesses its scope of applicability
through comparisons with numerical results from Monte Carlo simulations.
The paper ends with a discussion in Section \ref{sec_discussion}.
 
\section{Notations}
\label{sec_notations}

In the present paper, in order to lighten the notation system, the same notations are used 
to represent a random process (or random variable) and its realization. 
The elevation of a short-crested sea state is modeled as a random field, 
$\Pi(x,y,t)$, which depends on two horizontal cartesian coordinates, $x$, $y$, and a time variable, $t$. 
The free-surface elevation, $\Pi$, is measured vertically, relative to the mean water level.
The space coordinate system is attached to the reference frame of the mean flow 
(i.e. the reference frame where the mean Eulerian velocity of the fluid is zero).
In the subsequent developments, $\eta(t) = \Pi(x_0,y_0,t)$ and
$\de(t) =  {\rm d} \eta (t)/ {\rm d} t $
denote random processes
which respectively represent the second-order free-surface elevation and its time-derivative,
at a location $(x_0,y_0)$ fixed  in the reference frame of the mean flow.
A third random process, $\xi(t)$, will denote a third second-order wave variable 
which may have a dependency relationship with $\eta(t)$ and $\de(t)$.
The values taken by the these three random processes, at a given time, are denoted $\eta$, $\de$ and $\xi$.
In the present study, these random values 
will be first considered as non-conditioned (i.e. at a time fixed \textit{a priori}).
Then, the question of how the statistics of $\xi$ are changed 
when the observation time is conditioned to $\eta(t)$ upcrossing a given level, $\lc$, will be addressed.
When discussing the results, the predictions from the second-order wave model 
will be compared with those of the linear wave model.
For this purpose, the first-order component of a second-order quantity, $q$, will be denoted $q^{(1)}$.

In the main text, it will be always specified whether $\xi$ is non-conditioned 
or level-crossing conditioned.
In equations, the notation $\xi|\eta(t)\uparrow \lc$
will be used to denote the conditional value of $\xi(t)$, 
given that $\eta(t)$ upcrosses the level $\lc$. 
When approximating the probability density function of non-conditioned variables in terms of truncated Edgeworth series,
it is convenient to consider random variables in standardized form.
The wave variables considered in the present work ($\eta$, $\de$ and $\xi$), non-conditioned, 
all have a mean equal to zero.\footnote{\label{note_mean_zero}
This assumption does not limit the scope of the model to be developed below. 
If necessary, 
the random variables of the considered problem can be redefined beforehand, such that their means are equal to zero
}
The non-conditional standard deviation of a random variable $X$ will be denoted $\sigma_{X}$.
Hence, 
\begin{align}
\label{eq_rh}
\xv  & = \eta / \su \, , \\
\yv  & = \de / \sd \, , \\
\zv  & = \xi / \st \, ,
\end{align}
denote the standardized counterparts of the variables $\eta$, $\de$ and $\xi$.
Note that $\zv$ will not conserve, \textit{a priori}, its standardized character
when conditioned to level crossing.

\section{Edgeworth approximations}
\label{sec_edgeworth}

Edgeworth series express a probability density function in terms of a ``corrected'' Gaussian distribution
(although another kind of baseline distribution may be conceivable; see e.g. \cite{kolassa_2006}). 
Hence, this mathematical approach is particularly well suited to the study of ``weakly'' non-Gaussian processes.
The correction factor takes the form of a function series which multiplies the probability density function of a normal distribution.
The normal distribution takes the mean and the variance from the considered random variable 
(here the value of a random process at a given fixed time). 
The function series involves Hermite polynomials, which form an orthogonal set 
with respect to the density function of the standard normal distribution. 
Compared to the alike Gram-Charlier A series, 
Edgeworth series are arranged in such way that the order of a term in the series corresponds 
to the order of the correction it brings to the baseline normal distribution. 
Then, the error can be consistently controlled when the series is truncated at some order 
(see \cite{kolassa_2006} for more details).

In the context of ocean engineering,
Longuet-Higgins (1963) \cite{longuet_1963} 
used a truncated Edgeworth series to investigate the statistical properties of the sea elevation,
and compared the skewness predicted by the second-order potential flow theory with observed values.
Longuet-Higgins (1964) \cite{longuet_1964} also used a bivariate Edgeworth expansion 
to investigate the effect of nonlinearities on the zero-crossing frequency and the maxima distribution of a stationary random process.
Jensen (1995) \cite{jensen_1996} used a trivariate Edgeworth expansion, truncated to the leading order,
to approximate the conditional distribution of a second-order wave kinematic variable,
given that it is observed at a wave crest.
Jensen (2005) \cite{jensen_2005} also used a trivariate Edgeworth expansion 
to predict the average profile of a second-order wave,
given the magnitude of its crest height.

In the present study, Edgeworth's approximations are limited to non-Gaussian corrections of leading order,
involving only the third-order cumulants of the considered distributions
(see \S\ref{subsec_order_consistency} 
for further discussion on the order of Edgeworth's approximations).  
In \S\ref{subsec_ew_bivariate}, a bivariate Edgeworth approximation is provided for the non-conditioned joint distribution
of the free-surface elevation and its time derivative, respectively noted $\eta$ and $\de$.
In \S\ref{subsec_ew_trivariate}, a trivariate Edgeworth approximation is provided for the non-conditioned joint distribution 
of $\eta$, $\de$, and $\xi$, where $\xi$ is a third wave kinematic variable.

\subsection{Edgeworth approximation for the bivariate distribution of $\eta$ and $\de$}
\label{subsec_ew_bivariate}

The sea surface elevation, $\eta(t)$, is modeled as a time-differentiable stationary process,
which induces the non-correlation of $\eta$ and $\de$ at a given time:
\begin{equation}
\label{eq_eta_de_decor}
E \left[ \eta \de \right] = \frac{1}{2} E \left[ \frac{ {\rm d}  }{  {\rm d} t } \eta^2 \right] = 0 \, ,
\end{equation}
where $E$ is the expectation operator.
Taking advantage from the non-correlation of $\eta$ and $\de$, 
the Edgeworth approximation of their joint probability density function
may be expressed as (see e.g. \cite{longuet_1964}):
\begin{equation}
\label{eq_EW2D_dim}
\displaystyle \hat{f}_{\eta,\de}(\eta,\de)  = \frac{1}{\su\sd} \displaystyle \hat{f}_{\xv,\yv}(\eta/\su,\de/\sd) \, 
\end{equation}
where
\begin{equation}
\label{eq_EW2D}
\displaystyle \hat{f}_{\xv,\yv}(\xv,\yv) = \frac{\exp{-\frac{1}{2}(\xv^2+\yv^2)} }{2\pi} 
 \left\{ 1 + \left[ \lambda_{30} H_3(\xv) + 3 \lambda_{12} H_1(\xv) H_2(\yv) + \lambda_{03} H_3(\yv) \right]/6 \right\} \, .
\end{equation}
The functions $H_1$, $H_2$, $H_3$ are the three first probabilist's Hermite polynomials 
(their expressions are given in appendix \ref{subsec_hermite_1D}),
and $\lambda_{ab}$ are two-dimensional cumulants which may be expressed as
\begin{equation}
\label{eq_2D_cumulants}
\lambda_{ab} = E \left[ \xv^a \yv^b \right] \, , \ {\rm for} \ a+b \le 3 \, . 
\end{equation}
A ``hat'' symbol has been used in Eqs.~(\ref{eq_EW2D_dim}-\ref{eq_EW2D})  
to differentiate the Edgeworth approximations, 
$\hat{f}_{\eta,\de}$ and $\hat{f}_{\xv,\yv}$,
from the exact distributions, $f_{\eta,\de}$ and $f_{\xv,\yv}$.

\subsection{Edgeworth approximation for the trivariate distribution of $\eta$, $\de$, and a third variable $\xi$}
\label{subsec_ew_trivariate}

Taking into account the non correlation of $\eta$ and $\de$ (see Eq. \ref{eq_eta_de_decor}), 
the Edgeworth approximation of the probability density function
of the triad $(\eta,\de,\xi)$ may be expressed as (see \cite{jensen_1996}):
\begin{equation}
\label{eq_EW3D_dim}
\hat{f}_{\eta,\de,\xi}(\eta,\de,\xi) = \frac{1}{\su\sd\st} \hat{f}_{\xv,\yv,\zv}(\eta/\su,\de/\sd,\xi/\st) \, ,
\end{equation}
where
\begin{align}
\begin{split}
\label{eq_EW3D}
\hat{f}_{\xv,\yv,\zv} =  & \displaystyle \frac{1}{(2\pi)^{3/2}\sqrt{1-\rh^2-\rhp^2}} J_3 \\ 
 & \times \left[
1 + \frac{1}{6}\left( 
\lambda_{300} H_{300} 
+3\lambda_{201} H_{201}
+3\lambda_{120} H_{120}
+6\lambda_{111} H_{111} \right. \right. \\
& \left. 
\displaystyle +3\lambda_{102} H_{102}
+\lambda_{030} H_{030}
+3\lambda_{021} H_{021}
+3\lambda_{012} H_{012}
+\lambda_{003} H_{003} 
\right) \bigg] \, .
\end{split}
\end{align}
As in Eqs.~(\ref{eq_EW2D_dim}-\ref{eq_EW2D}), a ``hat'' symbol has been used in Eqs.~(\ref{eq_EW3D_dim}-\ref{eq_EW3D}) 
to differentiate the Edgeworth approximations, 
$\hat{f}_{\eta,\de,\xi}$ and $\hat{f}_{\xv,\yv,\zv}$,
from the exact distributions, $f_{\eta,\de,\xi}$ and $f_{\xv,\yv,\zv}$.
The function $J_3$ is the exponential term of the Gaussian distribution baseline:
\begin{equation}
\label{eq_J3}
J_3(\xv,\yv,\zv) = \exp \left\{ -\frac{1}{2} \frac{(1-\rhp^2)\xv^2+(1-\rh^2)\yv^2+\zv^2-2\rh\xv\zv-2\rhp\yv\zv+2\rh\rhp\xv\yv}{1-\rh^2-\rhp^2} \right\} \, .
\end{equation}
The coefficients 
\begin{align}
\label{eq_rh}
\rh  & = \frac{E [ \eta \xi ]}{\sigma_{\eta} \sigma_{\xi}} = E [ \xv \zv ]  \, , \\
\label{eq_rhp}
\rhp  & = \frac{E [ \de \xi ]}{\sigma_{\de} \sigma_{\xi}} = E [ \yv \zv ]  \, , 
\end{align}
are the non-conditional correlation coefficients of the pairs $(\eta,\xi)$ and $(\de,\xi)$.
The three-dimensional cumulants,
$\lambda_{abc}$, may be expressed as follows:
\begin{equation}
\label{eq_3D_cumulants}
\lambda_{abc} = E \left[ \xv^a \yv^b \zv^c \right] \, , \ {\rm for} \ a+b+c \le 3 \, . 
\end{equation}
Similarly to the pair $(\eta,\de)$, 
the processes $\eta^2(t)$ and $\de(t)$, considered at a given time, are uncorrelated with
\begin{equation}
\label{eq_eta2_de_uncorr}
E \left[ \eta^2 \de \right] = \frac{1}{3} E \left[ \frac{ {\rm d}  }{  {\rm d} t } \eta^3 \right] = 0 \, .
\end{equation}
\eq{eq_eta2_de_uncorr} translates into $\lambda_{210} = \lambda_{21} = 0$, 
result which has been used in Eqs.~(\ref{eq_EW2D}-\ref{eq_EW3D}).
The trivariate Hermite polynomials, $H_{abc}$, are given by:
\begin{equation}
\label{eq_hermite_gene}
H_{abc} (\xv,\yv,\zv) = \frac{(-1)^{a+b+c}}{J_3(\xv,\yv,\zv)} 
\frac{\partial^a}{\partial {\xv}^a} \frac{\partial^b}{\partial {\yv}^b} \frac{\partial^c}{\partial {\zv}^c} 
J_3 (\xv,\yv,\zv) \, .
\end{equation}
As calculations will show, 
explicit expressions for the third-order Hermite polynomials (with $a+b+c=3$), appearing in \eq{eq_EW3D},
are not required in the present study.

\subsection{Computation of cumulants}

Within the linear wave theory,
the realization of a random wave field  
can be approximated as the sum of independent Airy waves\footnote{ 
\eq{eq_ss_realisation_o1} is equivalent to the alternative parametrization
where the wave field is represented as $\Pi^{(1)}(x,y,t) \simeq \sum\limits_{n=1}^{N} \sum\limits_{q=1}^{Q} A_{nq} \cos \left[ \omega_n t - (k_n \cos \theta_q) x - (k_n \sin \theta_q) y + \phi_{nq} \right]$,
with $\phi_{nq}$ and $A_{nq}$ being random phases and amplitudes.
The relation between these two alternative parametrizations is readily obtained from trigonometric identities.
}
(see e.g. \cite{ochi_2005, holthuijsen_2007})
\begin{align}
\label{eq_ss_realisation_o1}
\begin{split}
\Pi^{(1)}(x,y,t) \simeq \sum\limits_{n=1}^{N} \sum\limits_{q=1}^{Q}  
 & a_{nq} \cos \left[ \omega_n t - (k_n \cos \theta_q) x - (k_n \sin \theta_q) y \right] \\
 & + b_{nq} \sin \left[ \omega_n t - (k_n \cos \theta_q) x - (k_n \sin \theta_q) y \right] \, ,
\end{split}
\end{align}
where the angular frequencies and directions, $\omega_n$, $\theta_q$, 
account for the discretization of the two-dimensional wave spectrum.
The wave amplitudes $a_{nq}$ and $b_{nq}$, are independent random variables,
which follow centered normal distributions of variance 
\begin{equation}
\label{eq_elementary_variance}
{\sigma_{nq}}^2 = G(\omega_n,\theta_q) \Delta \omega_n \Delta \theta_q  \, ,
\end{equation}
where $G(\omega, \theta)$ is the one-sided variance density spectrum 
(defined for $\omega > 0$ and $\theta \in ]-\pi,\pi]$), 
and $\Delta \omega_n$,  $\Delta \theta_q$ 
are the sizes of frequency and direction discretization intervals.
A sea state realization is fully defined by the realization of the wave amplitudes $a_{nq}$ and $b_{nq}$.
The wave numbers, $k_n$, are related to the wave frequencies, $\omega_n$, through the dispersion relation
\begin{equation}
\label{eq_disp_rel}
\omega_n^2 = g k_n \tanh k_n h \, ,
\end{equation}
where $g$ is the acceleration due to gravity,
and $h$ is the water depth. 

In the present approach, the space coordinate system $(x,y,z)$ is Eulerian. 
A Lagrangian-type
wave model may have been an interesting alternative, as this kind of model, in its first-order approximation, 
has been shown to account for interesting nonlinear wave features (see e.g. \cite{gjosund_2003, fouques_2006, guerin_2019, lindgren_2021}), 
which are missed by the Eulerian linear model 
(e.g. the steepening of crests and flattening of troughs). 
The Lagrangian approach leads to a model which is nonlinear 
when expressed in an Eulerian coordinate system, even when it is restricted to the first order
(which explains why it can render some nonlinear features).
This is due to the nonlinearities introduced by the ``mapping'' between Eulerian and Lagrangian coordinates.
This mapping makes the crossing conditioning problem less readily tractable than in Eulerian approaches.
Within the first-order Lagrangian wave model,
Lindgren \& Lindgren \cite{lindgren_2011} addressed this matter
by making use of a multivariate form of Rice's formula, 
evaluating the resulting distributions through Monte Carlo integration.
The present study focuses on the level-crossing conditioning in the second-order Eulerian wave model.

Let us consider the second-order values of the free-surface elevation, $\eta(t) = \Pi(x_0,y_0,t)$, its time derivative $\de(t)$, 
and a third wave variable $\xi(t)$, all measured at a fixed horizontal location $(x_0,y_0)$.
In the present section the random values of the processes are considered non-conditioned,
i.e. considered at a time, $t_0$, given \textit{a priori}.
As the sea state is assumed to be stationary and homogeneous,
the spacetime location $x_0 = y_0 = t_0 = 0$ may be chosen without loss of generality.
The random wave amplitudes $a_{nq}$ and $b_{nq}$ are collected in a random vector U as follows:
\begin{align}
\label{eq_U_vector}
\begin{split}
U & = [u_i]_{i=1,...,2NQ} \\
    & = [a_{11},...,a_{1Q},a_{21},...,a_{2Q},...,a_{NQ},b_{11},...,b_{1Q},b_{21},...,b_{2Q},...,b_{NQ}]^{\intercal} \, .
\end{split}
\end{align}
Then, the second-order values of the wave variables, 
at this specific spacetime position, may be expressed as
\begin{align}
\label{eq_eta_tf_qtf_1}
\eta  & = \alpha_{i} u_i + \alpha_{ij} u_i u_j  \, , \\
\label{eq_eta_tf_qtf_2}
\de & = \beta_{i} u_i + \beta_{ij} u_i u_j  \, , \\
\label{eq_eta_tf_qtf_3}
\xi    & = \gamma_{i} u_i + \gamma_{ij} u_i u_j  \, ,
\end{align}
where the Einstein summation convention is used.
The first and second terms on the right-hand sides of Eqs. (\ref{eq_eta_tf_qtf_1}-\ref{eq_eta_tf_qtf_2}-\ref{eq_eta_tf_qtf_3})
are respectively the first-order and second-order contributions of the wave model.
The coefficients $\alpha_{i}$, $\beta_{i}$, $\gamma_{i}$ are the linear transfer functions of the respective wave variables, in discretized form, 
and the coefficients $\alpha_{ij}$, $\beta_{ij}$, $\gamma_{ij}$ are their quadratic transfer functions (QTFs), also in discretized form.
For instance the linear transfer coefficients $\alpha_i$ 
are readily obtained from Eqs. (\ref{eq_ss_realisation_o1}-\ref{eq_U_vector}),
setting $x=y=t=0$:
\begin{equation}
\alpha_i = \left\{\begin{array}{l}    1 \ \ {\rm for} \ i=1,...,NQ \\
                                                    0 \ \ {\rm for} \ i=NQ+1,...,2NQ
              \end{array}\right.
\end{equation}
The exhaustive list of linear transfer functions and quadratic transfer functions
of wave variables obtained from Stokes' perturbative approach are not reproduced here.
In the general case of short-crested 
sea states in finite-depth waters (as considered in the present study),
the material necessary to express these functions can be found in a number of publications
(see e.g. \cite{sharma_1980,pingxing_1994,dalzell_1999,molin_2002}).

Starting from Eqs. (\ref{eq_eta_tf_qtf_1}-\ref{eq_eta_tf_qtf_2}-\ref{eq_eta_tf_qtf_3}), the joint cumulants of $\eta$, $\de$ and $\xi$, 
may be expressed in terms of linear coefficients, quadratic coefficients 
and Airy wave variances.
The first order cumulants are given by
\begin{align}
\label{eq_means_expression_1}
K_{100} = & E[\eta] = \alpha_{ii} V_i  \, , \\
\label{eq_means_expression_2}
K_{010} = & E[\de] = \beta_{ii} V_i   \, , \\
\label{eq_means_expression_3}
K_{001} = & E[\xi] = \gamma_{ii} V_i \, ,
\end{align}
where $V_i = E[{u_i}^2]$ is the variance of $i^{\rm th}$ Airy wave amplitude,
which is given by Eqs. (\ref{eq_elementary_variance}-\ref{eq_U_vector}).
In the present study, the QTF of $\eta$ is defined such that $\eta$ is measured relative to the mean water level;
i.e. $K_{100} = E[\eta] = 0$.
The variable $\de$, as the time derivative of $\eta$, has also a zero mean, 
$K_{010} = 0$.
As for the third variable, in the examples considered in Section \ref{sec_illust_examples}, 
its mean is also zero\footnote{See footnote \ref{note_mean_zero}.}, $K_{001} = 0$.
The other cumulants, up to the third order, are given by 
\begin{equation}
\label{eq_cum3D_gene}
K_{abc} = E[\eta^a\de^b\xi^c] \, ,  \ a+b+c \le 3 \, .
\end{equation}
Longuet-Higgins \cite{longuet_1963} proposed a general procedure to compute cumulants of arbitrary order,
introducing the concept of `irreducible terms'.
In the present case, cumulants may be also directly computed by substituting 
Eqs. (\ref{eq_eta_tf_qtf_1}-\ref{eq_eta_tf_qtf_2}-\ref{eq_eta_tf_qtf_3}) into \eq{eq_cum3D_gene}, yielding for example
\begin{align}
\label{eq_cumulants_O2}
K_{101} & = E[\eta \zeta] = 
\alpha_i \gamma_i V_i 
+ 2 \alpha_{ij} \gamma_{ij} V_i V_j  \, , \\
\label{eq_cumulants_O3}
K_{111} & = E[\eta\de\xi]  = 
2 \alpha_i \beta_j \gamma_{ij} V_i V_j 
+ 2 \alpha_i \gamma_j \beta_{ij} V_i V_j
+ 2 \beta_i \gamma_j \alpha_{ij} V_i V_j
+ 8 \alpha_{ij} \beta_{ik} \gamma_{jk} V_i  V_j V_k \, .
\end{align}
In \eq{eq_cumulants_O2} the quadratic transfer coefficients have been assumed to be given in symmetric form
with $\alpha_{ij}=\alpha_{ji}$, $\beta_{ij}=\beta_{ji}$, $\gamma_{ij}=\gamma_{ji}$.
The expression of other second-order and third-order cumulants
may be easily deduced from 
Eqs. (\ref{eq_cumulants_O2}-\ref{eq_cumulants_O3}).
Then, the cumulants in standardized form, as appearing in Eqs. (\ref{eq_EW2D}-\ref{eq_EW3D}), 
may be computed as 
(for the cumulants appearing in Eq.~\ref{eq_EW2D}, 
note that $\lambda_{ab} = \lambda_{ab0}$)
\begin{equation}
\lambda_{abc} = \frac{K_{abc}}{\sqrt{ {K_{200}}^a {K_{020}}^b {K_{002}}^c }} \, .
\end{equation}
In Eqs. (\ref{eq_cumulants_O2}-\ref{eq_cumulants_O3}),
the sums of terms involving $E[{u_i}^4]$ and $E[{u_i}^6]$ have been neglected,
as they become respectively negligible compared to the sums of terms $V_iV_j$ and $V_iV_jV_k$,
when the number of considered wave harmonics becomes large ($NQ \rightarrow + \infty$);
i.e. when the wave frequency discretization step 
(along with the wave direction discretization step for continuous directional distributions) tends to zero.

\subsection{Order consistency of Edgeworth approximations}
\label{subsec_order_consistency}

In Eqs. (\ref{eq_EW2D}-\ref{eq_EW3D}), 
the third order cumulants bring first-order modifications to the normal distribution,
in the sense that 
\begin{equation}
\lambda_{abc} = O\left(\sqrt{\sum\limits_{i} V_i}\right) = O(\eta) \, , \ {\rm for} \ a+b+c=3.
\end{equation}
The terms of the type $2 \alpha_{ij} \gamma_{ij} V_i V_j$ and $8 \alpha_{ij} \beta_{ik} \gamma_{jk} V_i  V_j V_k$, 
as appearing in Eqs. (\ref{eq_cumulants_O2}-\ref{eq_cumulants_O3}), 
respectively correspond to $O(\eta^2)$
and $O(\eta^3)$ corrections to the second- and third-order standardized cumulants.
When considering order consistency, these terms may be neglected, 
since the Edgeworth approximations, as expressed in Eqs. (\ref{eq_EW2D}-\ref{eq_EW3D}), 
are truncated to the leading order. 
In the present study these higher-order terms are retained.
These terms will be useful in Section \ref{sec_illust_examples}, 
where the vertical component of the fluid velocity at the mean water level, $\vz$, 
is one of the wave variables considered as the third random variable of the problem.
In the linear wave theory, ${\de}^{(1)} = {\vz}^{(1)}$, resulting in ${\rhp}^{(1)} = 1$.
Then, if the $O(\eta^2)$ terms in the computation of $\rhp$ were to be neglected, 
the Edgeworth approximation as written in \eq{eq_EW3D} would break down,
since the baseline normal distribution would become degenerate.
Apart from this specific point, 
it has been checked that the decision of retaining or neglecting higher-order terms does not significantly change the results
to be presented in Section \ref{sec_illust_examples}.

\section{Applying Rice's formulae to Edgeworth approximations}
\label{sect_rice_ew}

\subsection{Upcrossing frequency}
\label{subsec_upcross_freq}

The upcrossing frequency of the free-surface elevation, $\eta$, at the level $\lc$ 
is given by Rice's formula \cite{rice_1944, rice_1945}:
\begin{align}
\displaystyle \nuz(\lc) & = \int_0^{+ \infty} \yvi' f_{\eta,\de}(\lc,\yvi') \ {\rm d}\yvi'  \nonumber \\
\label{eq_rice_freq}
                                 & = \frac{\sd}{\su}\int_0^{+ \infty} \yvi f_{\xv,\yv}(\tilde{\lc},\yvi) \ {\rm d}\yvi  \, ,
\end{align}
where
\begin{equation}
\tilde{\lc} = \frac{\lc}{\su} \, ,
\end{equation}
$f_{\eta,\de}$ is the joint distribution of the pair $(\eta,\de)$, non-conditioned, 
and $f_{\xv,\yv}$ is its standardized version.
When $f_{\xv,\yv}$ is replaced by its Edgeworth approximation $\hat{f}_{\xv,\yv}$ (see Eq. \ref{eq_EW2D}),
the upcrossing frequency becomes 
\begin{align}
\displaystyle \hat{\nuz}(\lc) & = \frac{\sd}{\su}\int_0^{+ \infty} \yvi \ \hat{f}_{\xv,\yv}(\tilde{\lc},\yvi)  {\rm d}\yvi \nonumber \\
\label{eq_rice_freq}
                                 & = \frac{1}{2\pi} \frac{\sd}{\su} \exp\left\{-\frac{1}{2}\tilde{\lc}\,^2\right\} 
                                 \left[ \mathcal{I}_0 + \frac{1}{6} \left\{ \lambda_{30} \mathcal{I}_0 H_3(\tilde{\lc}) 
                                 + 3 \lambda_{12} \mathcal{I}_2 H_1(\tilde{\lc}) + \lambda_{03} \mathcal{I}_3 \right\} \right]\, ,
\end{align}
where the integrals $\mathcal{I}_p$ are defined as 
\begin{equation}
\displaystyle \mathcal{I}_p = \int_0^{+\infty} \yvi H_p(\yvi) \exp\left\{ - \frac{1}{2}\yvi^2 \right\} \ {\rm d} \yvi \, .
\end{equation}
Calculations yield $\mathcal{I}_0 = 1$, $\mathcal{I}_2 = 1$, $\mathcal{I}_3 = 0$, 
(integration by parts may be used to obtain the values of $\mathcal{I}_2$ and $\mathcal{I}_3$)
leading to
\begin{equation}
\label{eq_nucross_approx}
\displaystyle \hat{\nuz}(\lc) = \frac{1}{2\pi} \frac{\sd}{\su} \exp{-\frac{1}{2}\tilde{\lc}\,^2}
\left[ 1 + \frac{\lambda_{30}}{6} H_3(\tilde{\lc}) + \frac{\lambda_{12}}{2}H_1(\tilde{\lc}) \right] \, .
\end{equation}
An expression similar to \eq{eq_nucross_approx} had been already obtained by Longuet-Higgins 
(see Eq. 31 in \cite{longuet_1964}),
with an extra factor $2$ due to the counting of both upcrossings and downcrossings,
and a typo in the normalization factor 
(the factor ${K_{02}}^{1/2}$ appearing in Eq. 31 in \cite{longuet_1964} should be placed at the numerator).
For comparison, the upcrossing frequency predicted by the linear wave theory
may be expressed as
\begin{equation}
\label{eq_nucross_o1}
{\nuz}^{(1)}(\lc) = \frac{1}{2\pi} \frac{ \sigma_{\de^{(1)}} }{ \sigma_{\eta^{(1)}} } \exp{-\frac{1}{2}\left(\frac{\ell}{\sigma_{\eta^{(1)}}}\right)^2} \, .
\end{equation}
The factor appearing into square brackets, in \eq{eq_nucross_approx}, 
may be viewed as the leading-order non-Gaussian correction to the upcrossing frequency.

\subsection{Conditional distribution given upcrossing}
\label{subsec_edgeworth_condi_dist}

The conditional distribution of a wave variable $\xi$, given free-surface upcrossing, 
may be expressed in terms of generalized Rice's formula
\begin{align}
f_{\xi|\eta(t) \uparrow \lc} (\xi)  & = 
\frac{ \displaystyle \int_0^{+\infty} \yvi' f_{\eta,\de,\xi}(\lc,\yvi',\xi) \ {\rm d} \yvi' }{
\displaystyle \int_0^{+\infty} \yvi' f_{\eta,\de} (\lc,\yvi') \ {\rm d} \yvi' } \nonumber \\
\label{eq_rice_dist_gene}
& = \frac{1}{\st} \frac{ \displaystyle \int_0^{+\infty} \yvi f_{\xv\yv\zv}(\tilde{\lc},\yvi,\xi / \st) \ {\rm d} \yvi }{
\displaystyle \int_0^{+\infty} \yvi f_{\xv,\yv} (\tilde{\lc},\yvi) \ {\rm d} \yvi } \, .
\end{align}
When the distributions $f_{\xv\yv\zv}$ and $f_{\xv,\yv}$ are replaced by their Edgeworth approximations,
the following approximation is obtained:
\begin{equation}
\label{eq_ew_condDist_general}
\hat{f}_{\xi | \eta(t) \uparrow \lc} (\xi) 
= \frac{1}{\st} \frac{ \displaystyle \int_0^{+\infty} \yvi \hat{f}_{\xv\yv\zv}(\tilde{\lc},\yvi,\xi / \st) \ {\rm d} \yvi }{
\displaystyle \int_0^{+\infty} \yvi \hat{f}_{\xv,\yv} (\tilde{\lc},\yvi) \ {\rm d} \yvi } \, .
\end{equation}
The integral at the denominator is a normalization factor 
which has been already computed when considering the upcrossing frequency 
in Section \ref{subsec_upcross_freq}, 
and may be expressed as
(noting that $\lambda_{ab0} = \lambda_{ab}$):
\begin{equation}
\label{eq_norm_factor} 
\displaystyle \int_0^{+\infty} \yvi \hat{f}_{\xv,\yv} (\tilde{\lc},\yvi) \ {\rm d} \yvi
= 
\frac{1}{2\pi} e^{-\frac{1}{2}\tilde{\lc}\,^2} 
\left[ 1 + \frac{\lambda_{300}}{6} H_3(\tilde{\lc}) + \frac{\lambda_{120}}{2}H_1(\tilde{\lc}) \right] \, .
\end{equation}
The computation of the numerator of \eq{eq_ew_condDist_general} 
is a bit more involved. It may be decomposed as follows:
\begin{equation}
\label{eq_numerator_rice_ew}
\mathcal{F}(\tilde{\lc},\zv) = \displaystyle \int_0^{+\infty} \yvi \hat{f}_{\xv\yv\zv}(\tilde{\lc},\yvi,\zv) \ {\rm d} \yvi = 
\alpha_{000} G_1(\tilde{\lc},\zv) + \sum_{a+b+c=3} \alpha_{abc} \mathcal{I}_{abc}(\tilde{\lc},\zv)
\end{equation}
where $G_1$ is given by 
\begin{equation}
\label{eq_G1_general}
G_1(\tilde{\lc},\zv) = 
\int_0^{+\infty} \yvi J_3(\tilde{\lc},\yvi,\zv) \ {\rm d} \yvi \, ,
\end{equation}
the functions $\mathcal{I}_{abc}$ are given by
\begin{equation}
\label{eq_Iabc}
\mathcal{I}_{abc}(\tilde{\lc},\zv) = \int_0^{+\infty} \yvi H_{abc}(\tilde{\lc},\yvi,\zv) J_3(\tilde{\lc},\yvi,\zv) \ {\rm d} \yvi \, ,
\end{equation}
and $\alpha_{abc}$ (with $a+b+c=3$) are numerical factors whose expressions 
can be identified from \eq{eq_EW3D}.
The integrals $\mathcal{I}_{abc}$ may be analytically computed by using 
a combination of integrations by parts and Hermite polynomial algebra.
The calculations are a bit lengthy and the details are reported in Appendices 
A, B, C and D.
All calculations done, $\mathcal{F}$ may be expressed as follows
\begin{align}
\begin{split}
\label{eq_num_rice_ew}
\mathcal{F} (\tilde{\lc},\zv)  = & \Gamma_0(\tilde{\lc},\zv) G_0(\tilde{\lc},\zv) 
  + \Gamma_1(\tilde{\lc},\zv) G_1(\tilde{\lc},\zv) \\
 & + \left[ \beta_1 H_{100}(\tilde{\lc},0,\zv) 
  + \beta_2 H_{010}(\tilde{\lc},0,\zv) 
  + \beta_3 H_{001}(\tilde{\lc},0,\zv) \right] J_3(\tilde{\lc},0,\zv) \, ,
\end{split}
\end{align}
where $G_0$ is defined by
\begin{equation}
\label{eq_G0_general}
G_0(\tilde{\lc},\zv) = 
\int_0^{+\infty} J_3(\tilde{\lc},\yvi,\zv) \ {\rm d} \yvi \, .
\end{equation}
The expression of the first-order Hermite polynomials, $H_{100}$, $H_{010}$, $H_{001}$, 
are reported in Appendix A, Eqs.~(\ref{eq_hernite_h100}-\ref{eq_hernite_h010}-\ref{eq_hernite_h001}).
Note that $H_{100}(\xv,\yv,\zv) \ne H_1(\xv)$, $H_{010}(\xv,\yv,\zv) \ne H_1(\yv)$ and  $H_{001}(\xv,\yv,\zv) \ne H_1(\zv)$.
Closed-form expressions for the integrals $G_0$ and $G_1$ are reported in Appendix D.
$\Gamma_0$ and $\Gamma_1$ are polynomials which can be expressed in terms of other polynomials, $P_{ab}$ ($3 \ge a+b \ge 2$), as follows:
\begin{align}
\begin{split}
\label{eq_gammaZ}
\Gamma_0 = & \left( -3 r \alpha_{300} - s \alpha_{201}\right) P_{20}  \\
                      & + \left( -3 s \alpha_{003} - r \alpha_{102} + \alpha_{012} \right) P_{02} \\
                      & + \left( -2r \alpha_{201} - 2s \alpha_{102} + \alpha_{111} \right) P_{11} \, ,
\end{split}
\end{align}
 \begin{align}
\begin{split}
\Gamma_1 = \alpha_{300} P_{30} + \alpha_{201} P_{21} + \alpha_{102} P_{12} + \alpha_{003} P_{03} + \alpha_{000} \, ,
\end{split}
\end{align}
where $r$ and $s$ are numerical coefficients given by
\begin{align}
\label{eq_rc}
r =  & - \frac{\rh\rhp}{1-\rh^2} \, ,\\
\label{eq_sc}
s = & \frac{\rhp}{1-\rh^2} \, .
\end{align}
The polynomials $P_{ab}$ 
may be themselves conveniently expressed in terms of first-order polynomials,
\begin{align}
\label{eq_P11}
P_{11} = & S R + \frac{\rh}{1-\rh^2} \, , \\
\label{eq_P20}
P_{20} = & R^2 - \frac{1}{1-\rh^2} \, , \\
\label{eq_P02}
P_{02} = & S^2 - \frac{1}{1-\rh^2} \, , \\
\label{eq_P30}
P_{30} = & R \left( R^2-\frac{3}{1-\rh^2} \right) \, , \\
\label{eq_P03}
P_{03} = & S \left( S^2-\frac{3}{1-\rh^2} \right) \, , \\
\label{eq_P21}
P_{21} = & S \left( R^2-\frac{1}{1-\rh^2} \right) + 2 \frac{\rh}{1-\rh^2} R \, , \\
\label{eq_P12}
P_{12} = & R \left( S^2-\frac{1}{1-\rh^2} \right) + 2 \frac{\rh}{1-\rh^2} S  \, ,
\end{align}
where $R$ and $S$ are given by
\begin{align}
\label{eq_r_polynomials}
R(\xv,\zv) = & \frac{1}{1-\rh^2} \left( \xv -\rh \zv \right) \, ,  \\
\label{eq_s_polynomials}
S(\xv,\zv) = & \frac{1}{1-\rh^2} \left( -\rh \xv + \zv \right)  \, .
\end{align}
The terms $\beta_1$, $\beta_2$, $\beta_3$ are numerical coefficients which may be expressed as
\begin{align}
\beta_1 = & 3 r^2 \alpha_{300} + 2rs \alpha_{201} + s^2 \alpha_{102} - s \alpha_{111} + \alpha_{120}  \, , \\
\beta_2 =  & 2 (r^3 \alpha_{300} + r^2s\alpha_{201} +r s^2 \alpha_{102} + s^3 \alpha_{003}) 
                 - rs \alpha_{111} - s^2 \alpha_{012} + \alpha_{030}  \, , \\
\beta_3 = & r^2 \alpha_{201} + 2 rs \alpha_{102} + 3 s^2 \alpha_{003} - r \alpha_{111} - 2s \alpha_{012} + \alpha_{021} \, .
\end{align}

\section{Illustrative examples}
\label{sec_illust_examples}

The analytical approximations obtained in the previous section are now applied to a number of illustrative examples.
Both the upcrossing frequency and the conditional distribution of a kinematic variable are considered.

\subsection{Considered sea states}

\subsubsection{Shape of the two-dimensional wave spectrum}

Both long-crested and short-crested sea states have been considered in the present study.
The two-dimensional one-sided variance density spectrum $G(\omega,\theta)$ is assumed to be of the form
\begin{equation}
G(\omega, \theta) = D(\theta) \Omega(\omega) \, ,
\end{equation}
where
\begin{equation}
\int_{-\pi}^{\pi} D(\theta) \ {\rm d} \theta = 1 \, ,
\end{equation}
and $\Omega$ is the one-sided wave frequency spectrum.
In the considered examples, the sea states are assumed to have 
a JONSWAP frequency spectrum \cite{hasselmann_1973} 
with a peak enhancement parameter $\gamma=3.3$.
The normalization of the wave spectrum is set through the specification of the significant wave height, $H_s$, 
which is defined by:
\begin{equation}
\label{eq_jonswap_norm}
{H_s}^2 = 16 \int_{0}^{+\infty} \Omega(\omega) \ {\rm d} \omega \, .
\end{equation}
To avoid considering excessively long and short waves, 
the frequency spectrum is truncated at low and high frequencies:
$1\%$ of wave variance is discarded at low and high frequencies 
($2\%$ of wave variance is discarded in total).
Following this prescription, the low-frequency and high-frequency cutoffs lie respectively 
at $0.74 \omega_p$ and $3.0 \omega_p$, 
where $\omega_p$ is the peak angular frequency of the wave spectrum.
This truncation offers two benefits: 
(i) it makes the Monte Carlo simulation of sea state realizations (see Section \ref{subsec_MC_simus}, below) numerically less demanding;
(ii) it limits the issue of the poor convergence of the second-order perturbative solution, 
which arises when the interaction of two waves with an extreme wavelength ratio is considered
(see for example \cite{zhang_1993, zhang_1996, prevosto_2001}).
The normalization of the spectrum following \eq{eq_jonswap_norm} is performed before the truncation operation.

Two different wave direction distributions have been considered as case studies:
\begin{itemize}
\item Unidirectional sea state with a wave direction distribution 
\begin{equation}
\label{eq_D1_dist}
D_1(\theta)  = \delta(\theta),
\end{equation}
where $\delta$ is the Dirac delta function. 
Following the convention used for wave phases in \eq{eq_ss_realisation_o1}, 
\eq{eq_D1_dist} means that all the waves propagate in the direction of increasing $x$-coordinate.
\item Multidirectional sea state, with a spreading function given by
\begin{equation}
\label{eq_D2}
\begin{array}{cc|ll}
D_2(\theta) & = & (2/\pi) \cos^2 \theta & , \ {\rm for} \ \abs{\theta} < \pi/2 \\
  &  &  0 \,  & , \ {\rm for} \ \abs{\theta} > \pi/2 \, .
\end{array}
\end{equation}
\end{itemize}
It has been found that the choice of the direction distribution does not change qualitatively the results and their discussion.
Therefore, for the sake of conciseness, 
only results for the short-crested sea state ($D_2$) are reported in \S\ref{subsec_upcrossfreq}-\ref{subsec_cond_dist}.

\subsubsection{Magnitude of wave nonlinearities}

In the different examples to be reported in \S\ref{subsec_upcrossfreq}-\ref{subsec_cond_dist}, 
the effect of wave nonlinearity magnitude is investigated.
Two different sources of non-linearity are considered: the wave steepness and the water depth.
For a given sea state, a characteristic wave steepness may be defined as 
\begin{equation}
\kappa_p = k_p \frac{H_s}{2} \, ,
\end{equation}
where $k_p$ is the wave number corresponding to the peak frequency of the spectrum,
and $H_s$ is the significant wave height.
The water depth, $\dep$, is nondimensionalized as follows
\begin{equation}
\depa = k_p \dep \, .
\end{equation}
The level of nonlinearity brought by the finiteness of the water depth may be quantified by the function
\begin{equation}
\label{eq_dnl}
\dnl (\depa) = 
\frac{1}{2}  \left[ 3  \coth(\depa)^3 - \coth(\depa) \right] \, ,
\end{equation}
which corresponds to the amplification factor 
(relative to the case of infinite water depth) 
of the second-order correction 
for the free-surface elevation
of a regular wave of frequency $\omega_p$ 
(see for instance Eq. 3.60 in \cite{molin_2002}).
The function $\dnl$ has the following limits
\begin{equation}
\lim_{\depa \to 0} \dnl(\depa)  = +\infty \, , \  {\rm and} \ 
\lim_{\depa \to +\infty} \dnl(\depa)  = 1 \, ,
\end{equation}
 and it rapidly evolves in a quite narrow range of $\tilde{h}$ values,
 with $\dnl(2.2) \simeq 1.1$, $\dnl(1.2) \simeq 2.0$ and $\dnl(0.57) \simeq 10$.
 As a consequence, 
 the level of nonlinearity due to finite water depth 
 may be significantly larger for long waves (with frequencies smaller than $\omega_p$) 
 than for short waves (with frequencies larger than $\omega_p$).
Here, $\dnl(\depa)$ is adopted as a characteristic value.
 
 In the illustrative examples presented in Sections \ref{subsec_upcrossfreq}-\ref{subsec_cond_dist},
 the effects of increasing significant wave height
 and decreasing water depth 
 (both leading to an increase in wave nonlinearities) 
 are investigated separately.
 Seven different configurations of sea severity and water depth are considered as listed and labelled in Table \ref{table_cases}.
 The configuration series $\#$1-2-3-4
 will be used to illustrate the effect of increasing wave steepness in waters of infinite depth,
 while the series $\#$1-5-6-7 will be used to investigate the effect of decreasing water depth 
 for given values of $H_s$ and $T_p$.\footnote{
 Note that for given values of $H_s$ and $T_p$, a decrease in water depth, in addition to increasing $\dnl$, 
 also leads to an increase in wave steepness, 
 as reported in the last column of Table \ref{table_cases}.
 }


\begin{table}[h]
\centering
\begin{tabular}{|c|l|l|c|l|l|} 
 \hline
Configuration   & \multicolumn{1}{c|}{$T_p $} & \multicolumn{1}{c|}{$H_s$} & $\dnl$ & \multicolumn{1}{c|}{h} & \multicolumn{1}{c|}{$\kappa_p$}  \\ 
   \hline
$\#1$ & $10 \ {\rm s}$  & $1 \ {\rm m}$ & $1$ & \multicolumn{1}{c|}{$\infty$} & $\approx 2.0 \ 10^{-2}$    \\
 \hline
 $\#2$ & $10 \ {\rm s}$  & $2 \ {\rm m}$ & $1$ & \multicolumn{1}{c|}{$\infty$} & $\approx 4.0 \ 10^{-2}$    \\
 \hline
  $\#3$ & $10 \ {\rm s}$  & $4 \ {\rm m}$ & $1$ & \multicolumn{1}{c|}{$\infty$} & $\approx 8.0 \ 10^{-2}$    \\
 \hline
  $\#4$ & $10 \ {\rm s}$  & $6 \ {\rm m}$ & $1$ & \multicolumn{1}{c|}{$\infty$} & $\approx 0.12$    \\
 \hline
 $\#5$ & $10 \ {\rm s}$  & $1 \ {\rm m}$ & $2$ & $\approx 25 \ {\rm m}$ & $\approx 2.4 \ 10^{-2}$    \\
 \hline
 $\#6$ & $10 \ {\rm s}$  & $1 \ {\rm m}$ & $4$ & $\approx 14 \ {\rm m}$ & $\approx 3.0 \ 10^{-2}$    \\
 \hline
 $\#7$ & $10 \ {\rm s}$  & $1 \ {\rm m}$ & $6$ & $\approx 10 \ {\rm m}$ & $\approx 3.3 \ 10^{-2}$    \\
 \hline 
\end{tabular}
\caption{
List of the different configurations considered for illustrative purpose in Sections \ref{subsec_upcrossfreq}-\ref{subsec_cond_dist}. 
Each line of the table corresponds to a configuration.
The first column specifies a number which is used to identify the configuration. 
The second column specifies the assumed peak period, $T_p$.
The third column specifies the significant wave height, $H_s$.
The fourth column specifies the depth-related nonlinear factor $\dnl$ (see Eq. \ref{eq_dnl}).
The fifth column gives the value of the water depth corresponding 
to the assumed values of $T_p $ and $\dnl$.
The last column gives the characteristic wave steepness, $\kappa_p$, 
corresponding to the assumed values of $H_s$, $T_p$ and $\dnl$.
For all configurations, the wave direction distribution is assumed to be multidirectional, 
following \eq{eq_D2}.
}
\label{table_cases}
\end{table}


\subsection{Monte Carlo simulations}
\label{subsec_MC_simus}

In order to obtain a benchmark to be compared with the analytical approximations derived in Section \ref{sect_rice_ew},
Monte Carlo realizations of the sea states listed in Table \ref{table_cases} have been carried out. 
Both upcrossing frequencies and conditional distributions given upcrossing have been extracted 
from the Monte Carlo simulations.
A second-order wave code has been specifically developed for the present study.
Its implementation has been validated through a series of comparisons 
with a code independently developed by Prevosto in the framework of another project \cite{prevosto_2001}.
Comparisons have been carried out for long-crested/short-crested seas in finite/infinite water depths,
for the different kinematic variables considered in the present study 
(see Section \ref{subsubsec_considered_vars}).

In the case of short-crested seas, 
in order to make Monte Carlo simulations numerically less demanding, 
the number of discretization angles has been limited to 8.
Then, the same angle discretization has been used when computing the Edgeworth approximations to which 
Monte Carlo results are to be compared.
It has been checked that increasing the number of discretization angles does not change significantly the results. 

For each Monte Carlo realization, 
the wave spectrum is discretized in frequencies and directions 
(in the case of short-crested seas).
Following the discretized spectrum, the wave linear components 
(corresponding to the different discretization bins)
are first independently drawn 
with random phases and amplitudes (see e.g. paragraph 3.5 in \cite{holthuijsen_2007}, for details).
Then, the second-order corrections are computed and added to the linear components.
Hence, the resulting average second-order wave spectrum is different from 
the one specified as input 
(differences may be large in the low-frequency and high-frequency tails).
This is not an issue in the present study, 
as long as the same assumptions and inputs are adopted for the Monte Carlo simulations 
and the analytical model, which are to be compared below.

For each configuration listed in Table \ref{table_cases},
the number of simulated sea state realizations is $1.6 \times 10^4$,
each realization having a physical duration equal to $\simeq 341 \ T_p$.
Hence, for each configuration the total physical duration of simulated sea state is $\simeq 5.46 \times 10^6 \ T_p$.
This long duration yields a large amount of data which enable 
to probe low upcrossing frequencies 
(obtained for ``extreme'' crossing levels), 
as well as the tails of conditional distributions, given upcrossing.

\subsection{Upcrossing frequency}
 \label{subsec_upcrossfreq}
 

\def\scaleF{0.3}
\def\scaleL{0.2}
\def\scaleSpeLine{0.75}
\def\widthSpeLine{0.1}

\begin{figure}[t!]
\begin{center}
\begin{tabular}{lll} 
        &
        \includegraphics[width=\scaleF\textwidth]{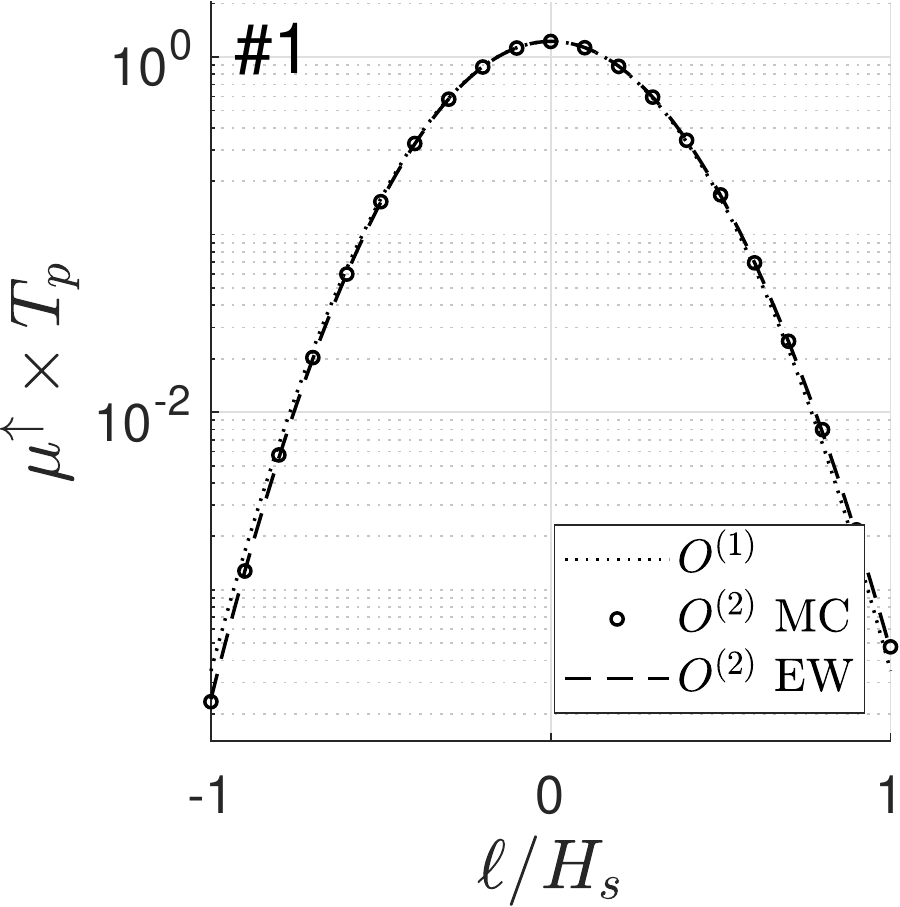} 
        &        
        \\
        \includegraphics[width=\scaleF\textwidth]{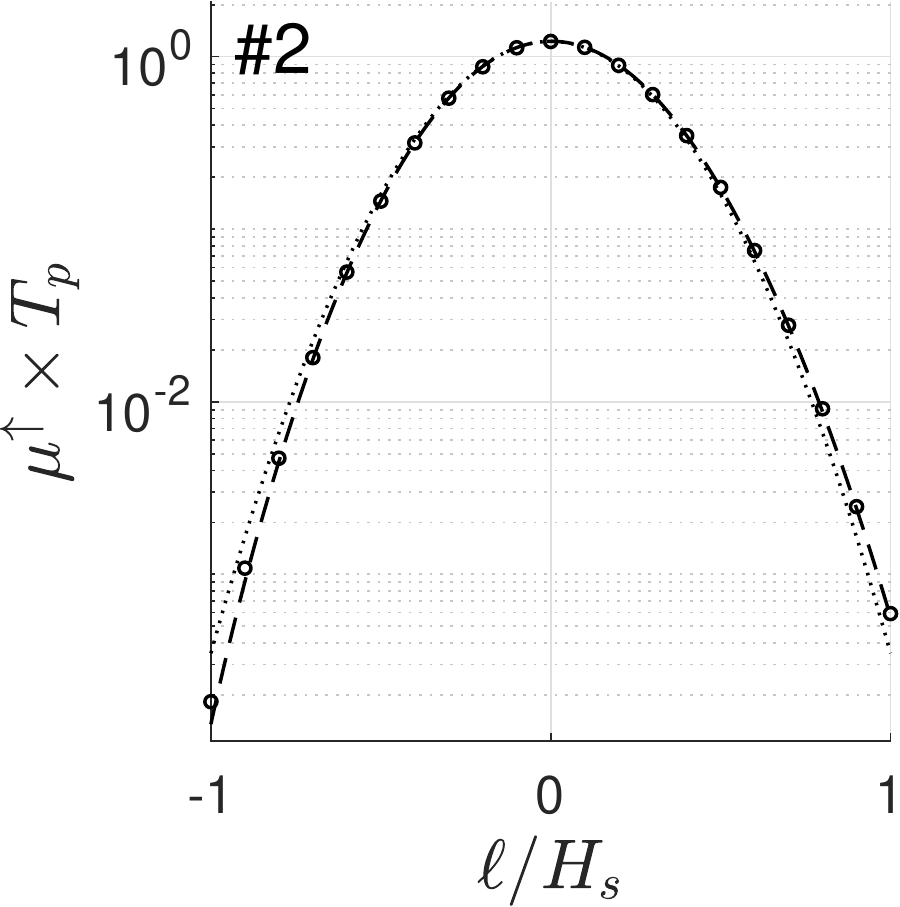} 
        & 
        \includegraphics[width=\scaleF\textwidth]{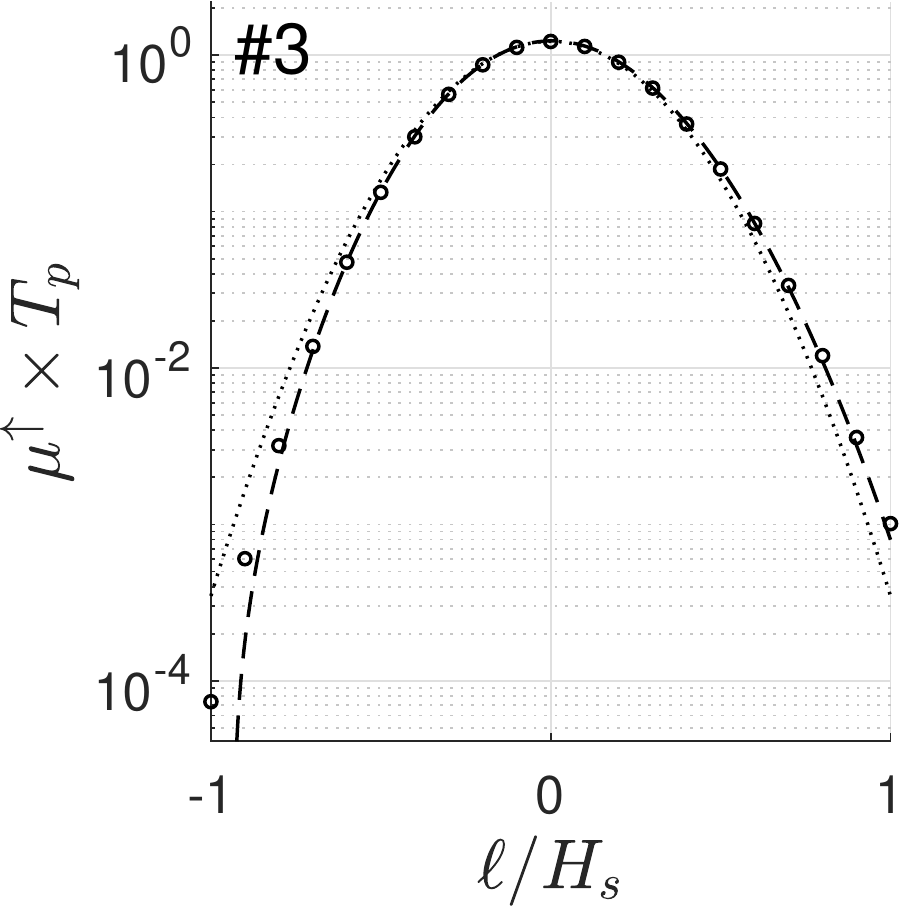} 
        &
        \includegraphics[width=\scaleF\textwidth]{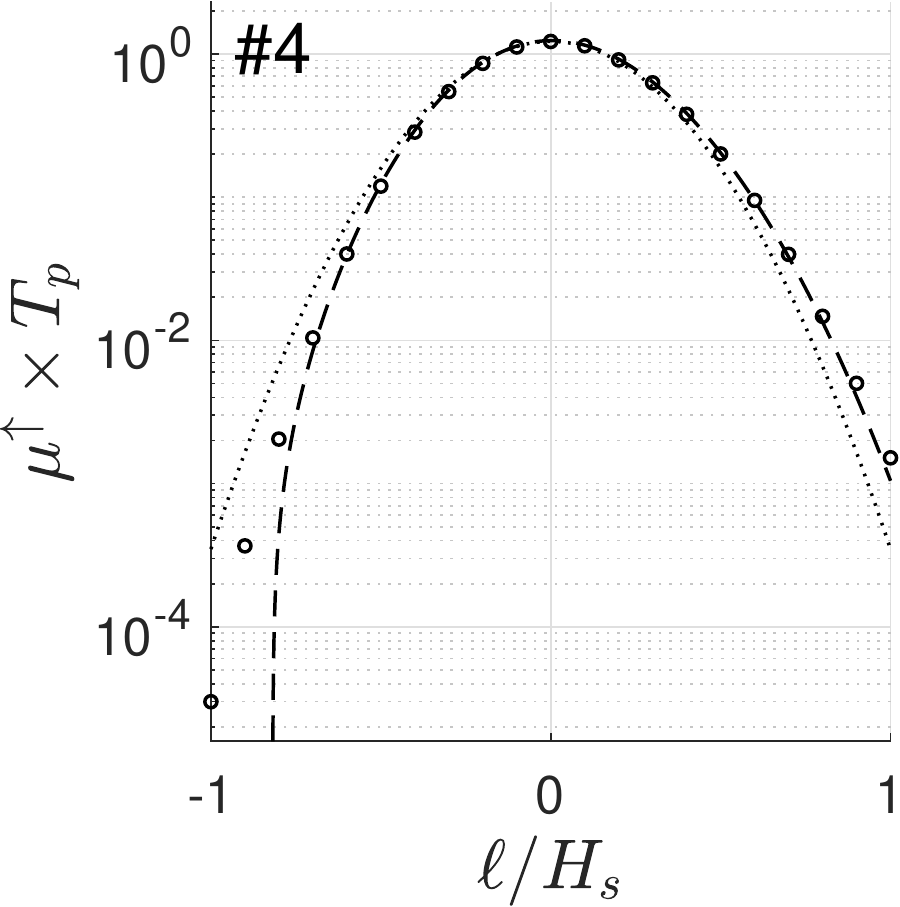} 
        \\
        \includegraphics[width=\scaleF\textwidth]{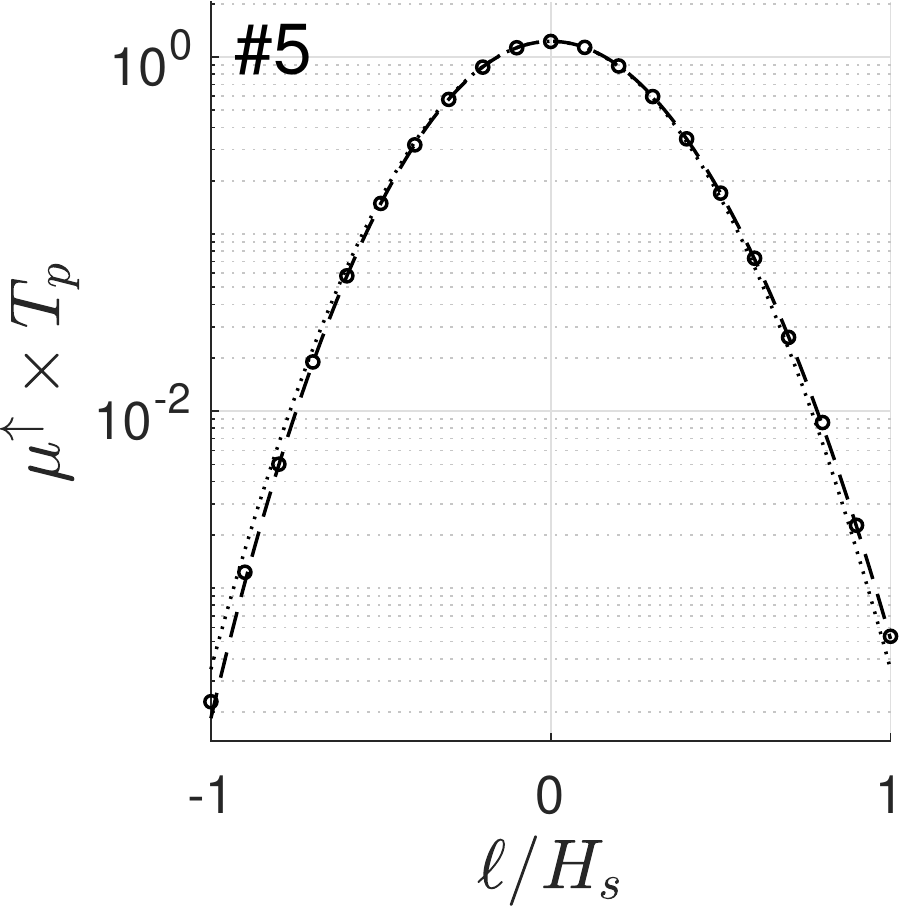} 
        & 
        \includegraphics[width=\scaleF\textwidth]{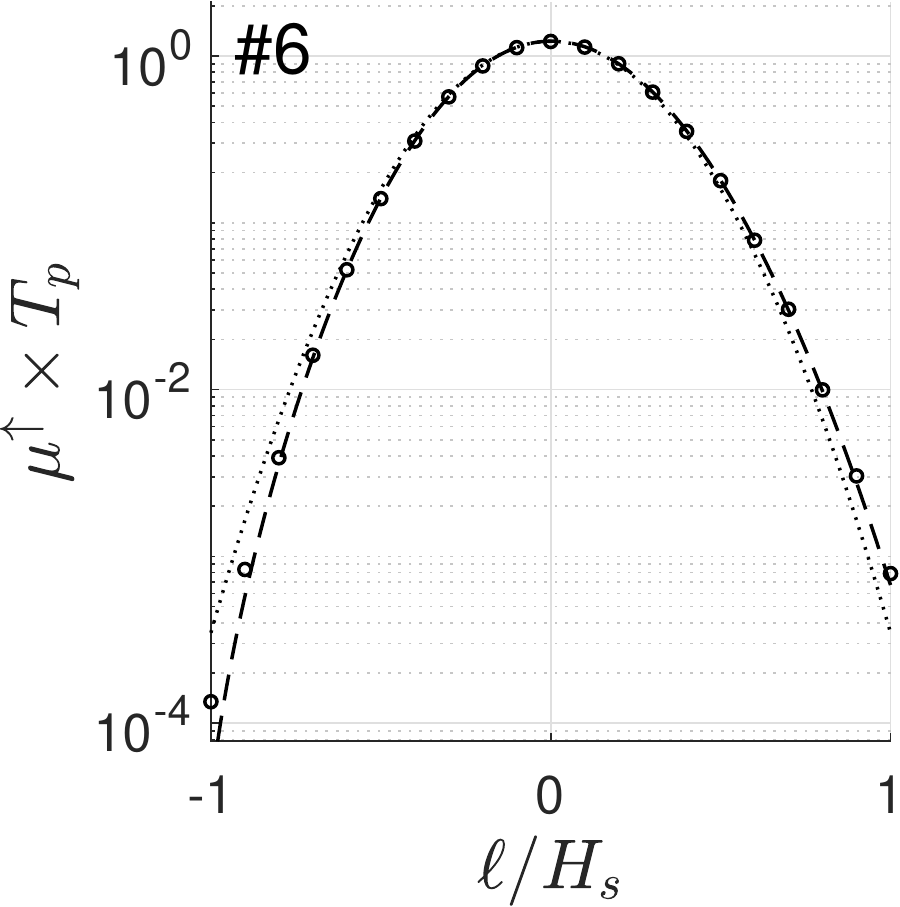} 
        &
        \includegraphics[width=\scaleF\textwidth]{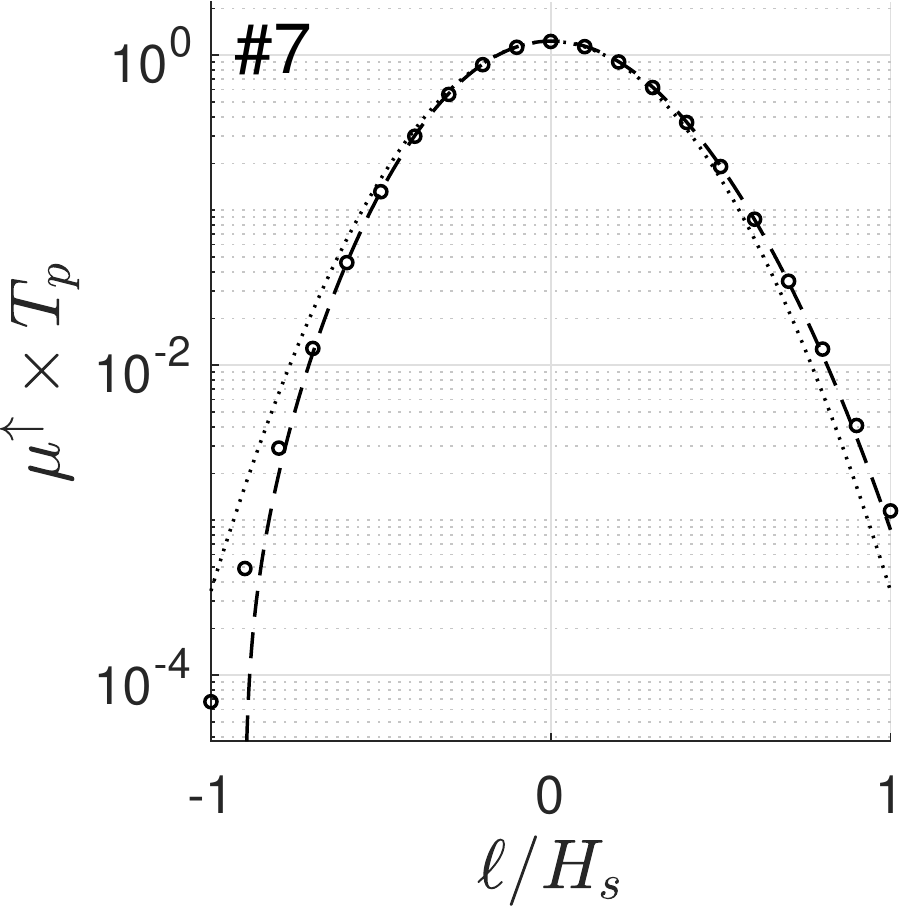} 
        \\        
\end{tabular}        
\end{center}
\caption{Upcrossing frequency as a function of the altitude of the level. The upcrossing frequency is shown for the linear wave model 
and the second-order model. In the case of the linear wave model, the upcossing frequency is obtained analytically 
from \eq{eq_nucross_o1} (dotted lines, labelled as `$O^{(1)}$').
For the second-order wave model, the upcrossing frequency obtained from Monte Carlo simulations 
(empty circles, labelled as `$O^{(2)}$ MC') 
is compared with the approximation obtained from Edgeworth's expansion (dashed line, labelled as `$O^{(2)}$ EW').}
\label{fig_example_vs_dists}
\end{figure}

 
 \fig{fig_example_vs_dists} shows the upcrossing frequency as a function of the altitude of the crossing level,
 for the different sea state configurations listed in Table \ref{table_cases}.
 The results are shown for both the linear and second-order wave models.
 Relative to the first-order prediction, 
 the second-order corrections increase (resp. decrease) the upcrossing frequency for positive (resp. negative) values of the altitude, $\ell$.
 This evolution can be well understood from the fact that
 second order corrections tend to increase the height of wave crests and decrease the depth of wave troughs, relative to the mean water level.

For the second-order wave model, the Monte Carlo estimates of the upcrossing frequency are compared 
with the analytical approximation derived from
Edgeworth's expansion (see Eq. \ref{eq_nucross_approx}). 
For weakly nonlinear configurations ($\# 1, 2, 5$) the analytical approximation is close
to numerical estimates in the range of considered crossing-level altitudes, $-1<\lc / H_s<1$.
When the magnitude of nonlinearities increase, 
either due to increasing $H_s$ ($\# 3, 4$)
or decreasing water depth ($\# 6, 7$),
the analytical approximation clearly deviates from the Monte Carlo estimates
in the regime of extreme events (i.e. for low upcrossing frequencies).
This is especially visible for $\lc/H_s$ close to $-1$, where a vertical asymptote begins to appear, in  \fig{fig_example_vs_dists}.
In fact, the Edgeworth approximation crosses zero in this region (hence the vertical asymptote in semi-logarithmic scale), 
and then becomes negative, which is unphysical. 
This is due to the Edgeworth correction factor being a cubic function of $\lc$ 
(see Eq. \ref{eq_nucross_approx}); 
then for $\lambda_{30} \ne 0$ ($\lambda_{30}$ being the skewness of the free-surface elevation, $\eta$) 
a range of negative values is inevitable 
for either extreme positive or extreme negative values of $\lc$ (depending on the sign of $\lambda_{30}$).
For all considered configurations, $\lambda_{30}$ is positive,
which explains that a cutoff appears in the Edgeworth approximation for negative crossing altitudes. 

In the range of positive crossing levels, the analytical approximation appears to slightly underestimate the second-order upcrossing frequency 
compared to Monte Carlo simulations.
This is particularly noticeable for the configuration $\#4$ and $\#7$, for $\lc / H_s$ close to $1$.
Overall, excluding the region where it 
plunges (in semi-logarithmic scale) to negative values,
the Edgeworth-type approximation appears to provide satisfactory predictions for the second-order upcrossing frequency
over a wide range of crossing-level altitudes.

 \subsection{Conditional distributions given upcrossing}
 \label{subsec_cond_dist}

To illustrate the second-order approximation obtained for the conditional distribution of a wave variable,
given free-surface upcrossing,
the derived model (see Eqs. \ref{eq_ew_condDist_general}-\ref{eq_norm_factor}-\ref{eq_num_rice_ew})
has been applied to three different kinematic variables, 
which are introduced in \S\ref{subsubsec_considered_vars}. 
In \S\ref{subsubsec_prob_densi} the analytical model is compared 
with Monte Carlo results, in terms of probability density function.
Then, in \S\ref{subsubsec_mvs_condi_dists}, the comparison
between the analytical and numerical results
focusses on the conditional mean, variance and skewness.
 
 \subsubsection{Considered kinematic variables}
 \label{subsubsec_considered_vars}

The three wave kinematic variables, to which free-surface upcrossing conditioning is applied, 
are the following: 
(i) $\vz$, the vertical component of the fluid velocity, at the mean water level,
(ii) $\sx$, the slope of the free surface along the direction $\theta= 0$ 
 (which is the average wave direction of the short-crested sea state considered in the present study; see Eq. \ref{eq_D2}),
 (iii) $\vx$, the horizontal component of the fluid velocity along the direction 
 $\theta=0$, at the mean water level.
When interested in the fluid velocity at the free surface, 
the values computed at the mean water level, $z=0$, may not be the best proxy. 
On the other hand, considering the velocity potential directly at $z=\eta$
may yield unrealistic fluid kinematics, especially in the crests.
This is related to the fact that short waves ``ride over'' long waves, in irregular seas. 
This issue has been extensively investigated within the linear wave model through comparisons
with experimental measurements
(see \cite{wheeler_1970,chakrabarti_1971,rodenbusch_1986,horng_1991,gudmestad_1993,xu_1995}
for instance).
A simple approach to build a second-order proxy for fluid kinematics at the free surface, 
could be to modify the considered QTFs 
by adding the second-order contribution coming from the Taylor expansion of fluid variables, 
from $z=0$ to $z=\eta$.
Other approaches may be inspired by solutions already proposed for the linear theory,
such as ``stretching''-type models 
(see \cite{wheeler_1970,chakrabarti_1971,rodenbusch_1986}
for instance).
Alternatively, an approach based on the ``hybrid wave model'', proposed by Zhang et al. (1996) \cite{zhang_1996}, 
may be considered.
Depending on the details of the chosen approach, 
the resulting proxy for the fluid kinematics at the free surface may differ.
Assessing these different approaches  
would require comparisons with experimental data,
which is beyond the scope of the present study.
Here, for the sake of simplicity, the fluid velocity components, $\vz$ and $\vx$, 
are derived from the second-order velocity potential 
considered at the mean water level, $z=0$.

Within the linear wave model,
the free-surface upcrossing conditioning 
has the following effects on the three wave variables listed above:
 \begin{enumerate}
 \item The vertical component of the fluid velocity at the mean water level 
 is equal to the time derivative of the free-surface elevation, $\vzL = \deL$.
 Then, as the random process $\eL(t)$ is Gaussian,
$\vzL | \eL(t) \uparrow\lc$ follows a Rayleigh distribution
whose mode is equal to the non-conditional standard deviation of $\deL$, $\sigma_{\deL}$.
Hence, the conditional distribution of $\vzL$, 
given upcrossing, does not depend on the actual crossing level, $\lc$. 
 \item 
 The free-surface slope, $\sxL(t) = {\Pi_{,x}}^{(1)}(x_0,y_0,t)$, non-conditioned, 
 is correlated to $\deL$, but it does not depend on $\eL$.
 Then, the resulting probability density function, given upcrossing, 
 may be expressed as the convolution 
 of a normal distribution with a Rayleigh distribution.
 Or alternatively, $\sxL | \eL(t)  \uparrow\lc$ may be viewed as resulting 
 from the sum of two independent random variables,
 one being Gaussian and the other being Rayleigh-distributed (see e.g. \cite{aberg_2008, hascoet_2021}).
 The normal distribution is centered with a variance equal to $[1-(\rhpL)^2] {\sigma^2_{\sxL}}$,
 where $\rhpL$ is the non-conditional correlation coefficient between $\deL$ and $\sxL$,
 and $\sigma_{\sxL}$ is the non-conditional standard deviation of $\sxL$.
 The Rayleigh distribution has a mode equal to $\abs{\rhpL} \sigma_{\sxL}$,
 and the related random variable is to be added to (resp. subtracted from) 
 the Gaussian variable
 if $\rhpL > 0$ (resp. $\rhpL < 0$) -- 
 see appendix A.1.2 in \cite{hascoet_2021} for more details.
 Similarly to $\vzL$, the conditional distribution of $\sxL$, 
 given upcrossing, does not depend on the actual crossing level, $\lc$.
 \item 
 The horizontal component of the fluid velocity,
\begin{align}
\label{eq_vx_realisation_o1}
\begin{split}
\vxL(t) = \sum\limits_{n=1}^{N} g k_n/\omega_n \sum\limits_{q=1}^{Q}  
 &  \cos \theta_q \cdot \left\{ a_{nq} \cos \left[ \omega_n t - (k_n \cos \theta_q) x_0 - (k_n \sin \theta_q) y_0 \right] \right. \\
 & \left. + b_{nq} \sin \left[ \omega_n t - (k_n \cos \theta_q) x_0 - (k_n \sin \theta_q) y_0 \right] \right\} \, ,
\end{split}
\end{align}
 non-conditioned, is correlated to $\eL$, 
 but it does not depend on $\deL$ (see Section 3 in \cite{hascoet_2021}, for further details).
 Then, the conditional distribution of $\vxL$, given upcrossing, is Gaussian.
 The conditional mean is equal to 
 $(\sigma_{\vxL} / \sigma_{\eL}) \rhL \lc$,
 where $\sigma_{\vxL}$, $\sigma_{\eL}$ are the non-conditional standard deviations of $\vxL$, $\eL$,
 and $\rhL$ is the correlation coefficient between these two variables. 
 The conditional variance is given by $[1-(\rhL)^2] \sigma^2_{\vxL}$.
 \end{enumerate} 
Hence, within the linear wave model, the three selected variables have conditional distributions, given upcrossing, of different types.
Below in \S\ref{subsubsec_prob_densi}-\ref{subsubsec_mvs_condi_dists}, 
we address two different questions: (i) how nonlinearities modify these conditional distributions,
(ii) whether the Edgeworth-type analytical approximation is capable of rendering these modifications.

\subsubsection{Shape of the conditional probability density functions}
\label{subsubsec_prob_densi}


\begin{figure}[h!]
\begin{center}
\begin{tabular}{m{\scaleF\textwidth}m{\scaleF\textwidth}m{\scaleF\textwidth}} 
        \fbox{
        \includegraphics[width=\scaleL\textwidth,trim={1.33cm 2.2cm 0.8cm 3cm},clip]
        {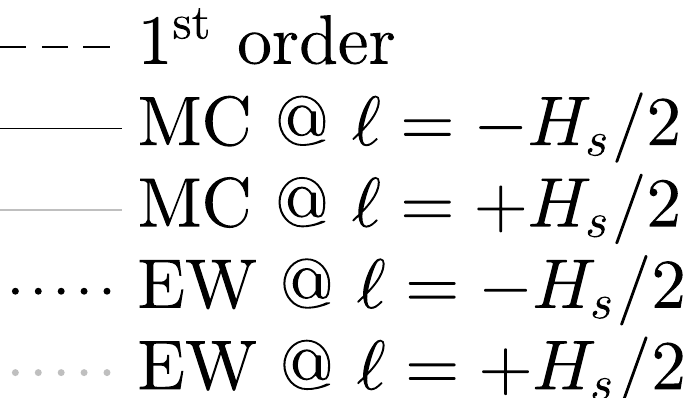}
        }
        &
        \includegraphics[width=\scaleF\textwidth]{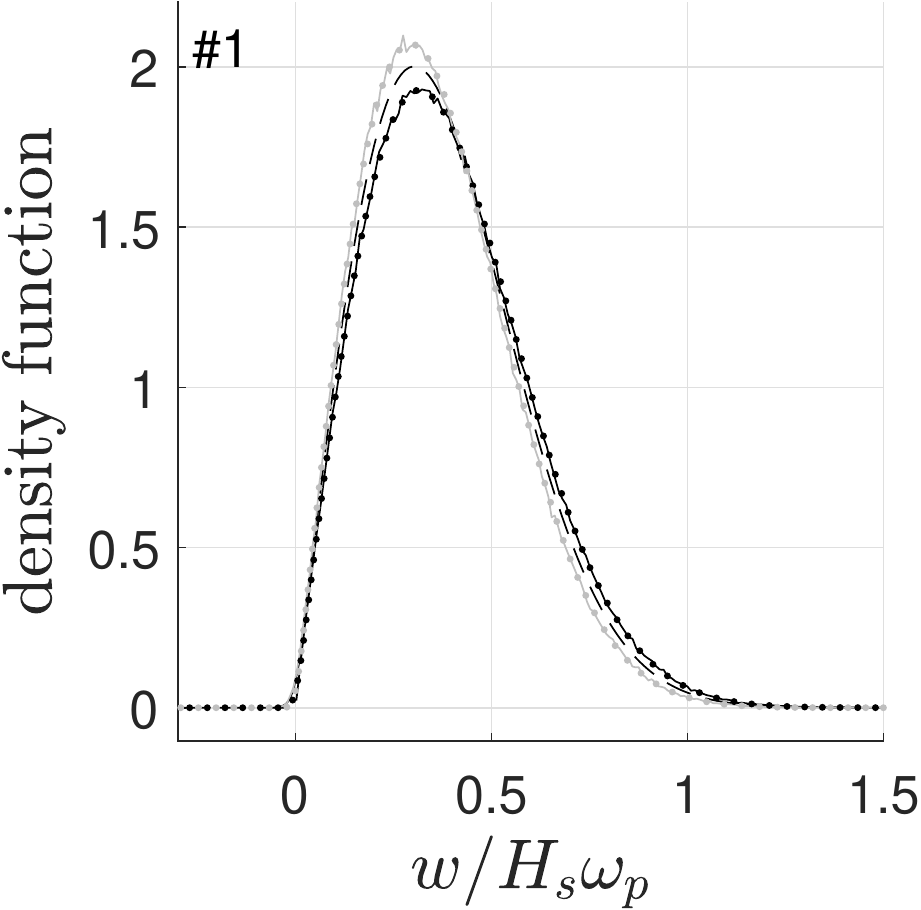} 
        &        
        \\
        \includegraphics[width=\scaleF\textwidth]{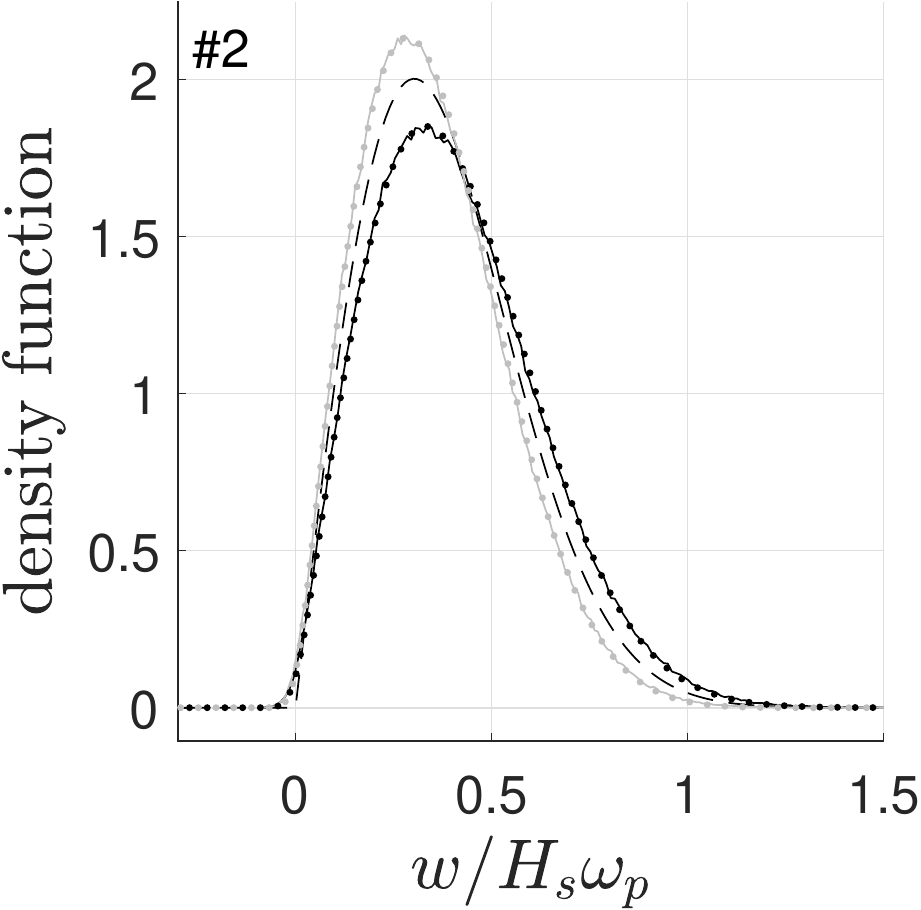} 
        & 
        \includegraphics[width=\scaleF\textwidth]{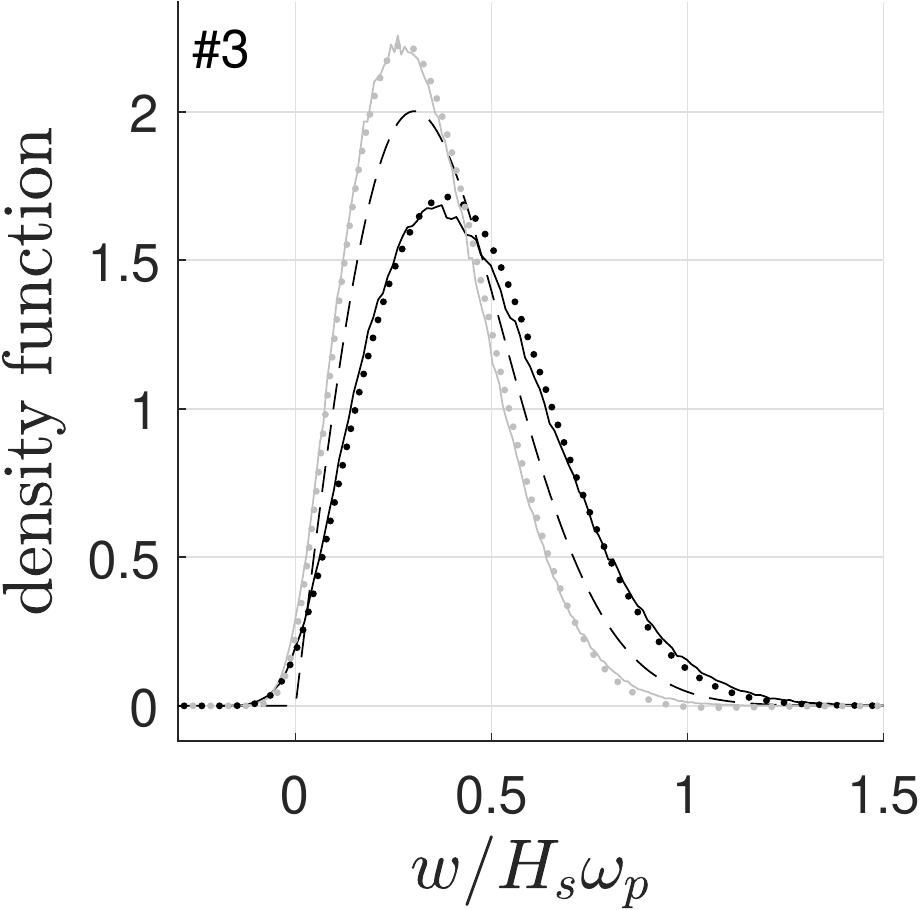} 
        &
        \includegraphics[width=\scaleF\textwidth]{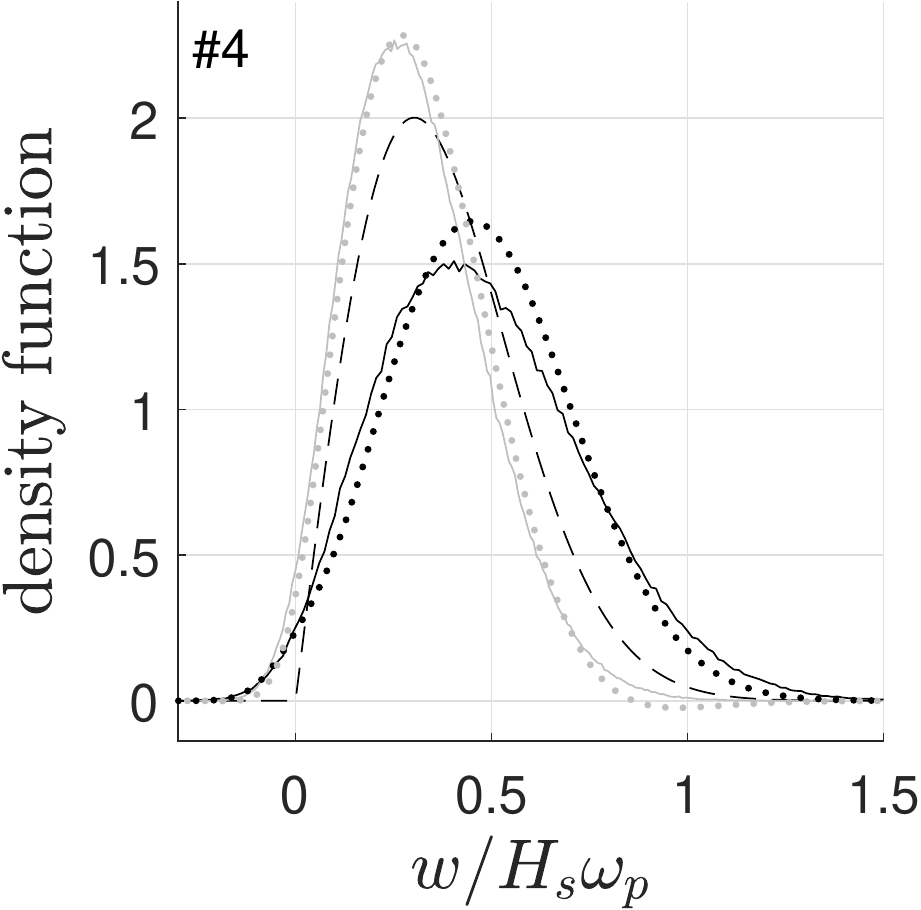} 
        \\
        \includegraphics[width=\scaleF\textwidth]{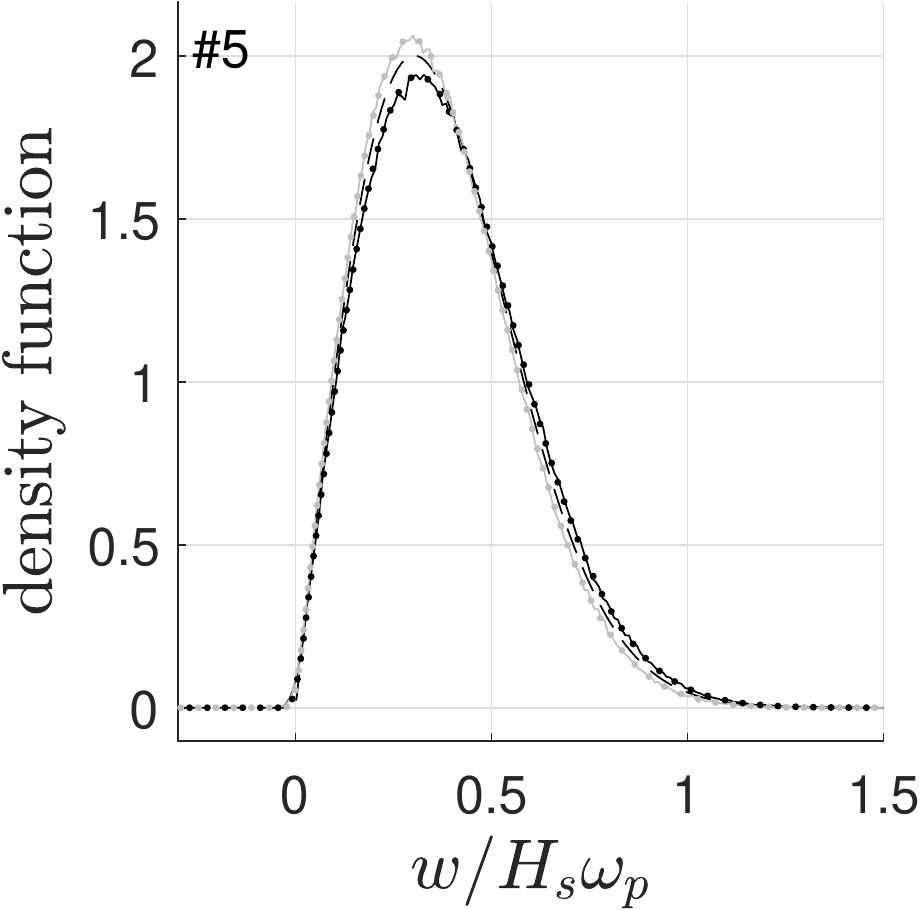} 
        & 
        \includegraphics[width=\scaleF\textwidth]{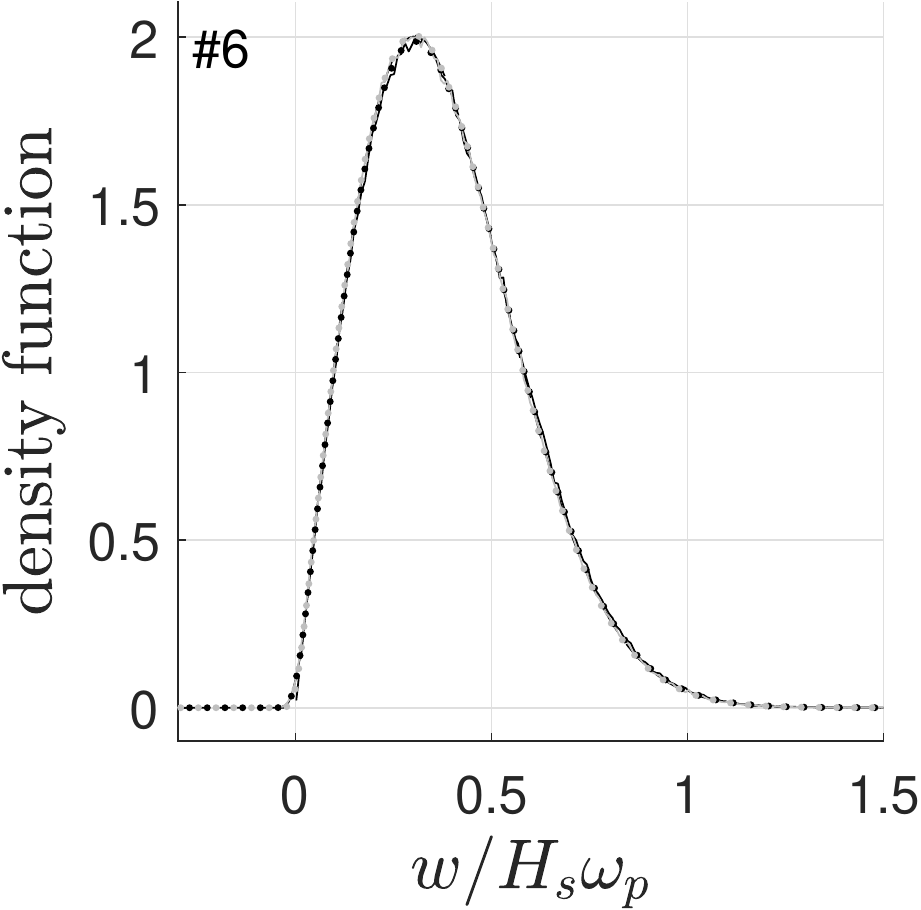} 
        &
        \includegraphics[width=\scaleF\textwidth]{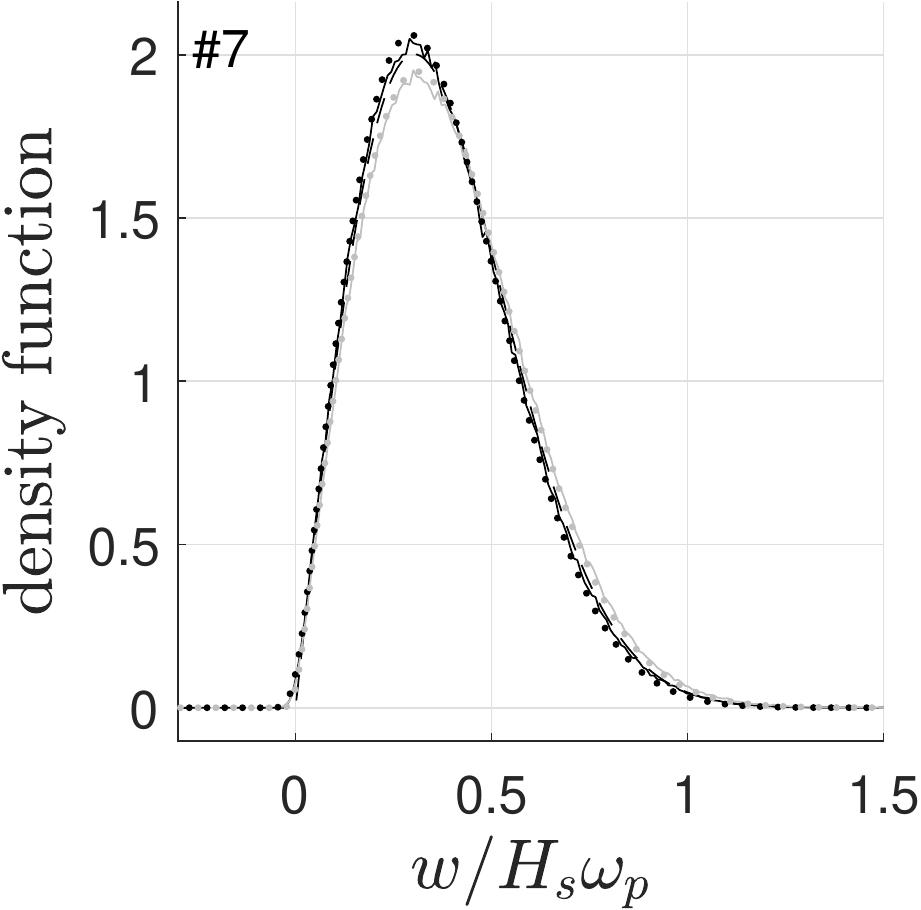} 
        \\        
\end{tabular}        
\end{center}
\caption{
Density function
of $\vz$, given free-surface upcrossing. 
The conditional distribution obtained from the linear wave model (a Rayleigh distribution) 
is shown as a dashed line (labelled as `1$^{\rm st}$ order' in the legend); it does not depend on the elevation of the crossing level. 
The conditional distribution obtained from the second-order wave model  is shown for two different crossing level elevations:
$\lc / H_s = -1/2$ (black)$;1/2$ (grey).
Results obtained from the Monte Carlo simulations (labelled as `MC') are shown as solid lines; 
the approximations based on Edgeworth expansions (labelled as `EW') are shown as dotted lines.
Each panel corresponds to a configuration listed in Table~\ref{table_cases}; 
the configuration number is indicated in the upper-left corner of each plot.}
\label{fig_uz_dists}
\end{figure}



\begin{figure}[h!]
\begin{center}
\begin{tabular}{m{\scaleF\textwidth}m{\scaleF\textwidth}m{\scaleF\textwidth}} 
        \fbox{
        \includegraphics[width=\scaleL\textwidth,trim={1.33cm 2.2cm 0.8cm 3cm},clip]
        {figLeg.pdf}
        }
        &
        \includegraphics[width=\scaleF\textwidth]{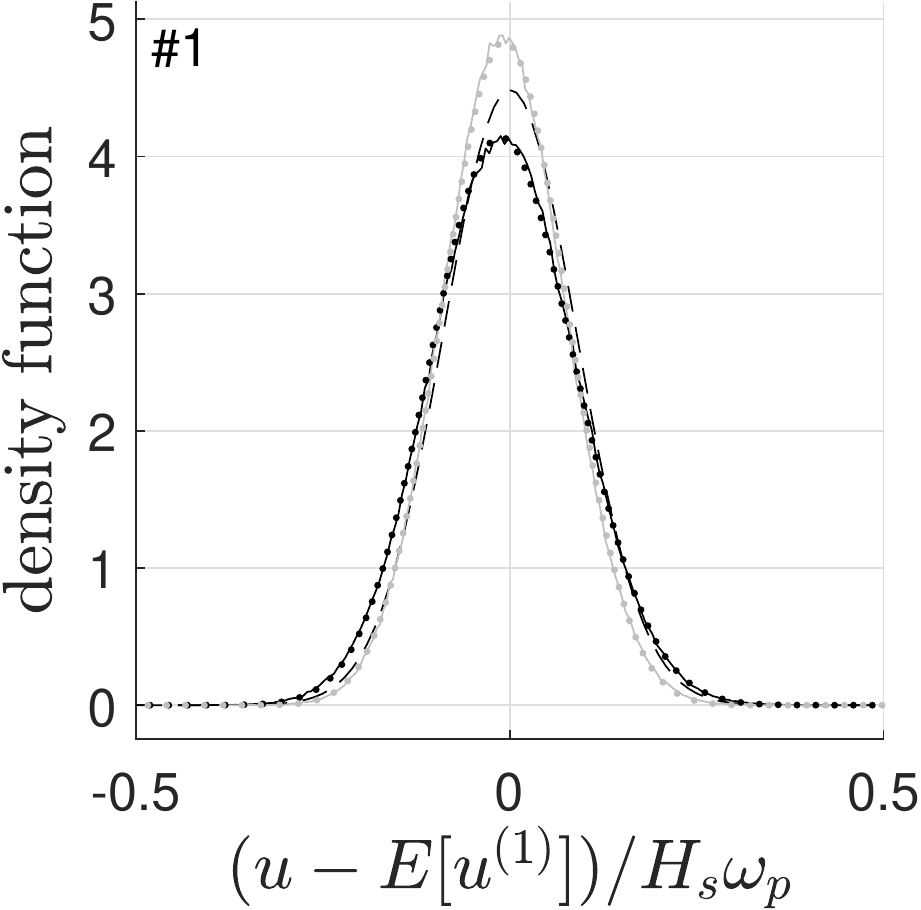} 
        &        
        \\
        \includegraphics[width=\scaleF\textwidth]{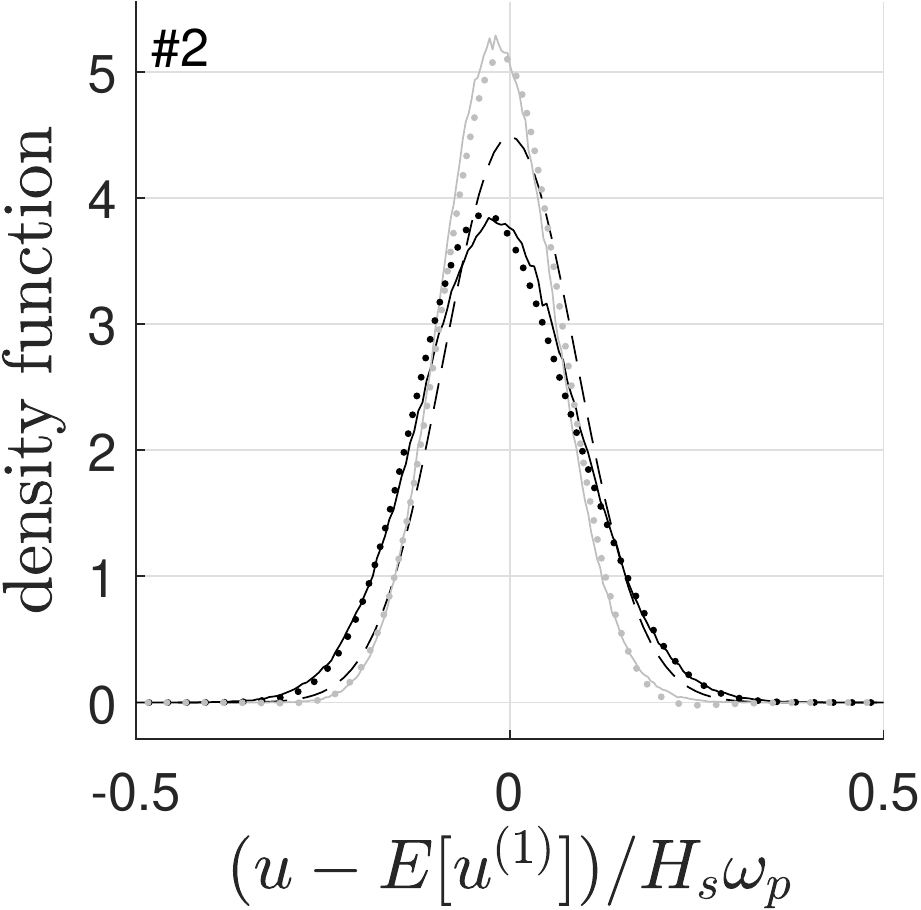} 
        & 
        \includegraphics[width=\scaleF\textwidth]{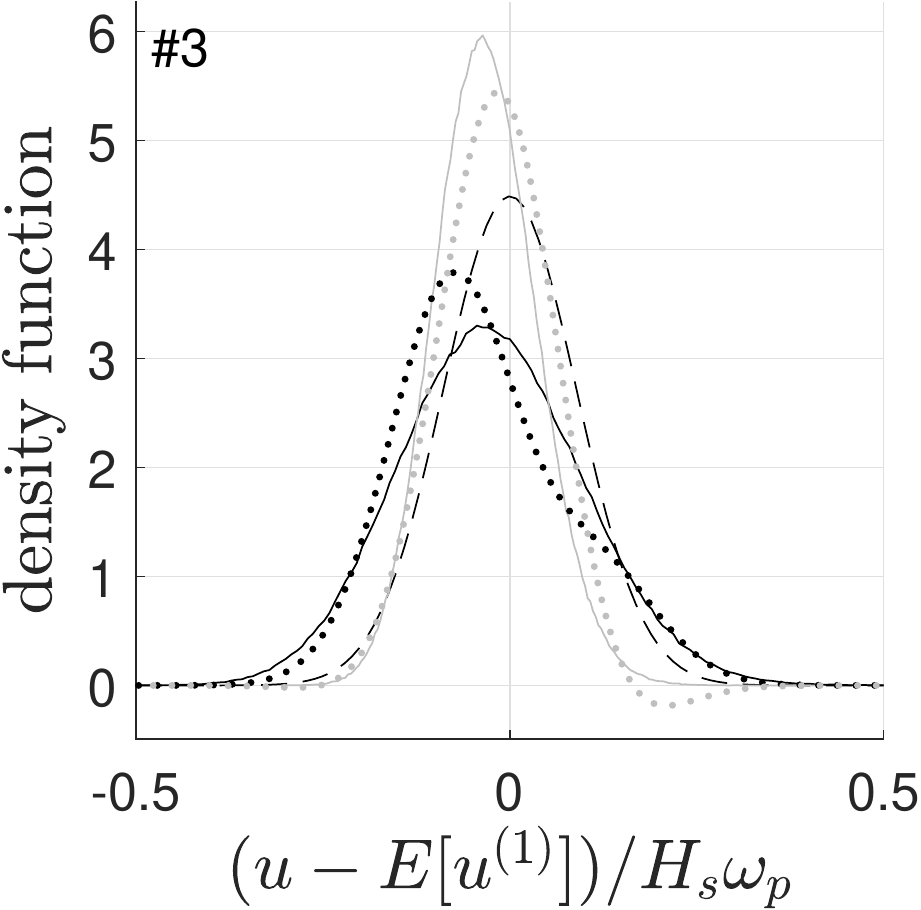} 
        &
        \includegraphics[width=\scaleF\textwidth]{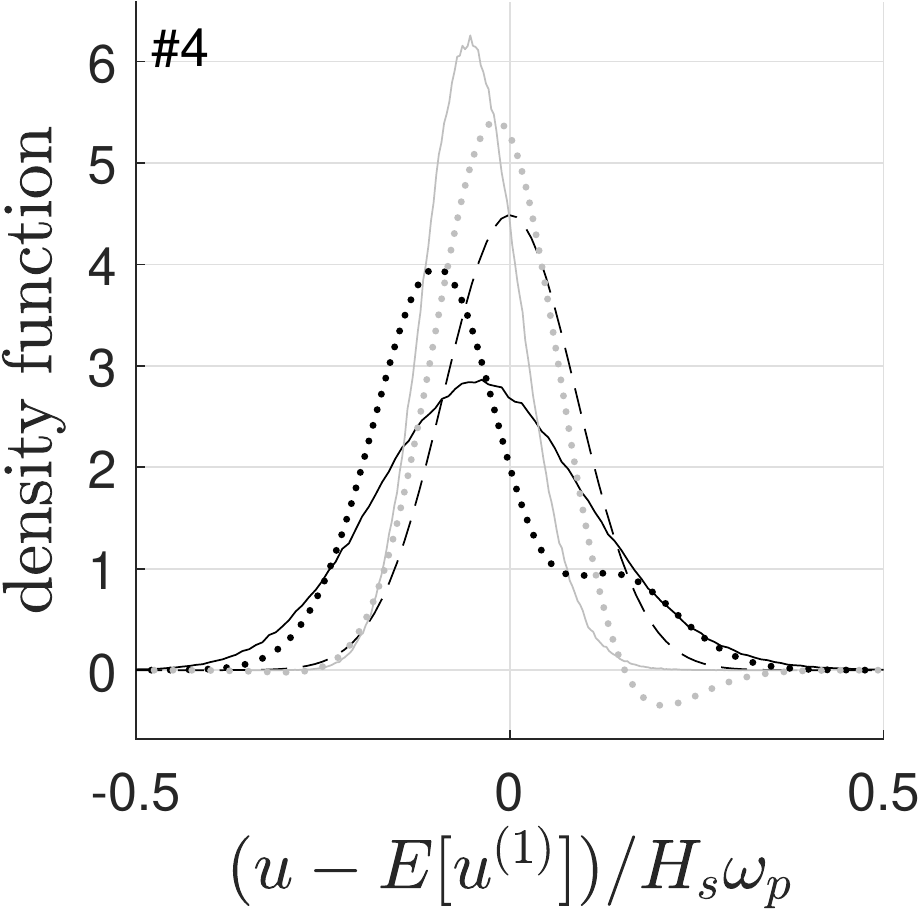} 
        \\
        \includegraphics[width=\scaleF\textwidth]{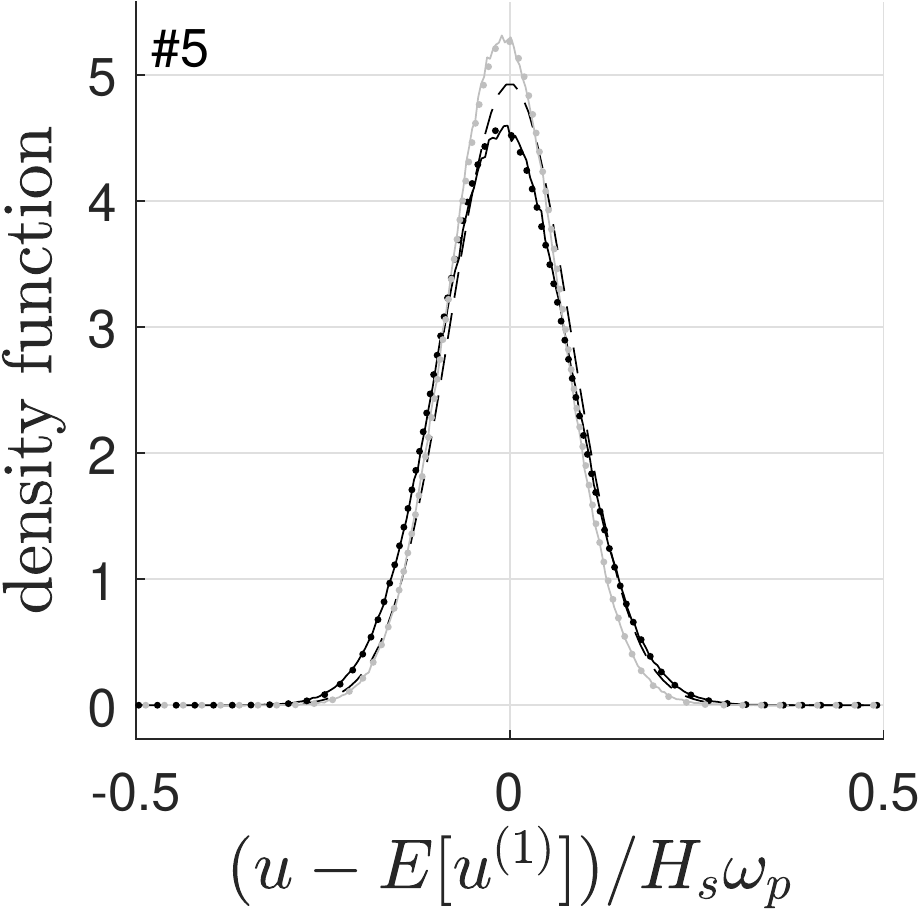} 
        & 
        \includegraphics[width=\scaleF\textwidth]{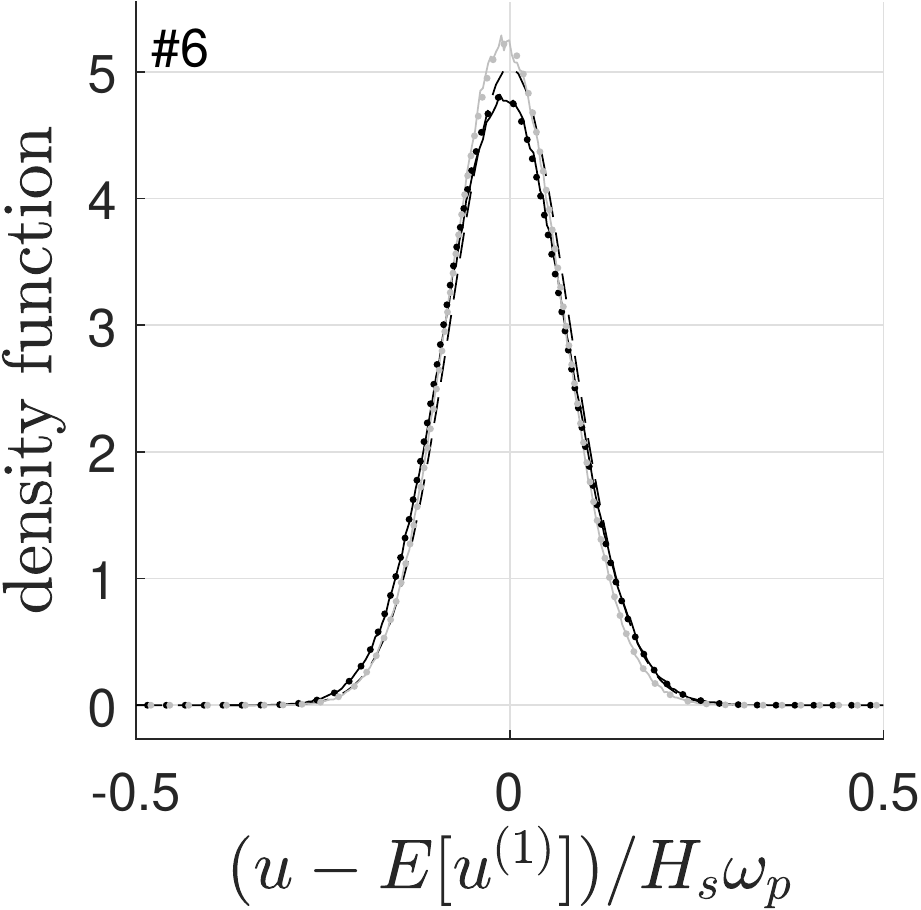} 
        &
        \includegraphics[width=\scaleF\textwidth]{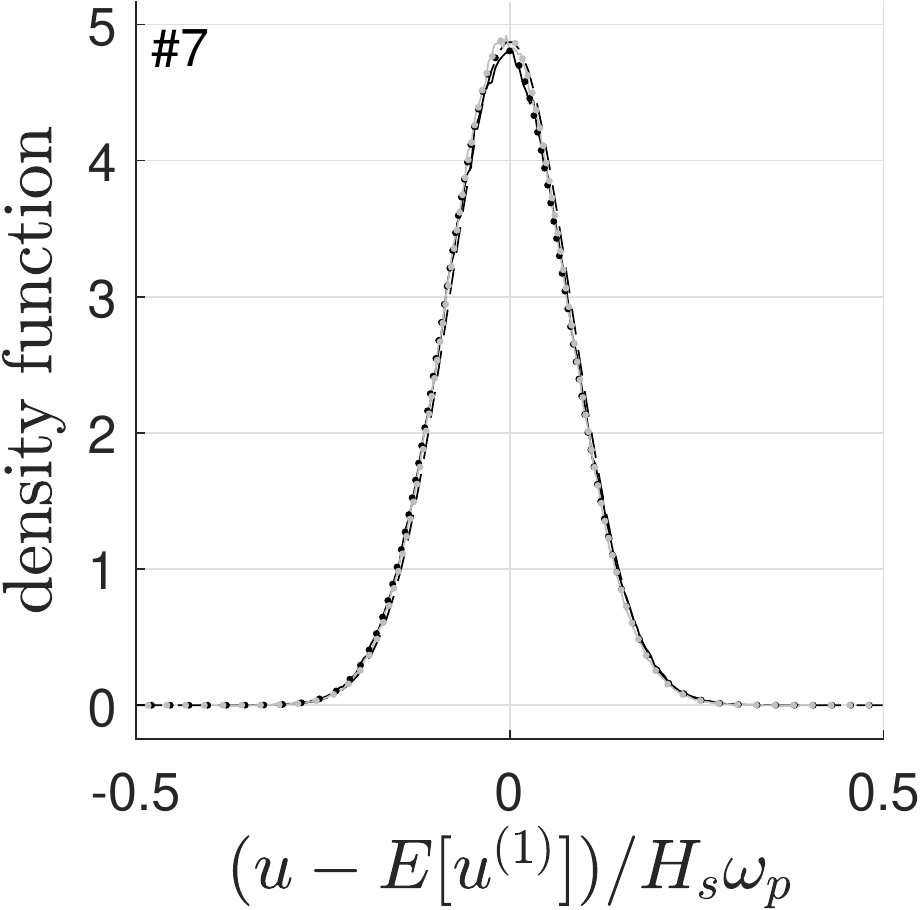} 
        \\        
\end{tabular}        
\end{center}
\caption{
Density function
of 
$\vx - E\left[ \vx^{(1)} \right]$, 
given  free-surface upcrossing.
   Here, $E\left[ \vxL\right]$ 
   denotes the conditional mean of $\vxL$, given that the process $\eL(t)$ upcrosses the level $\lc$.
    The way to read this figure is similar to the description given in the caption of \fig{fig_uz_dists}.}
\label{fig_ux_dists}
\end{figure}



\begin{figure}[h!]
\begin{center}
\begin{tabular}{m{\scaleF\textwidth}m{\scaleF\textwidth}m{\scaleF\textwidth}} 
        \fbox{
        \includegraphics[width=\scaleL\textwidth,trim={1.33cm 2.2cm 0.8cm 3cm},clip]
        {figLeg.pdf}
        }
        &
        \includegraphics[width=\scaleF\textwidth]{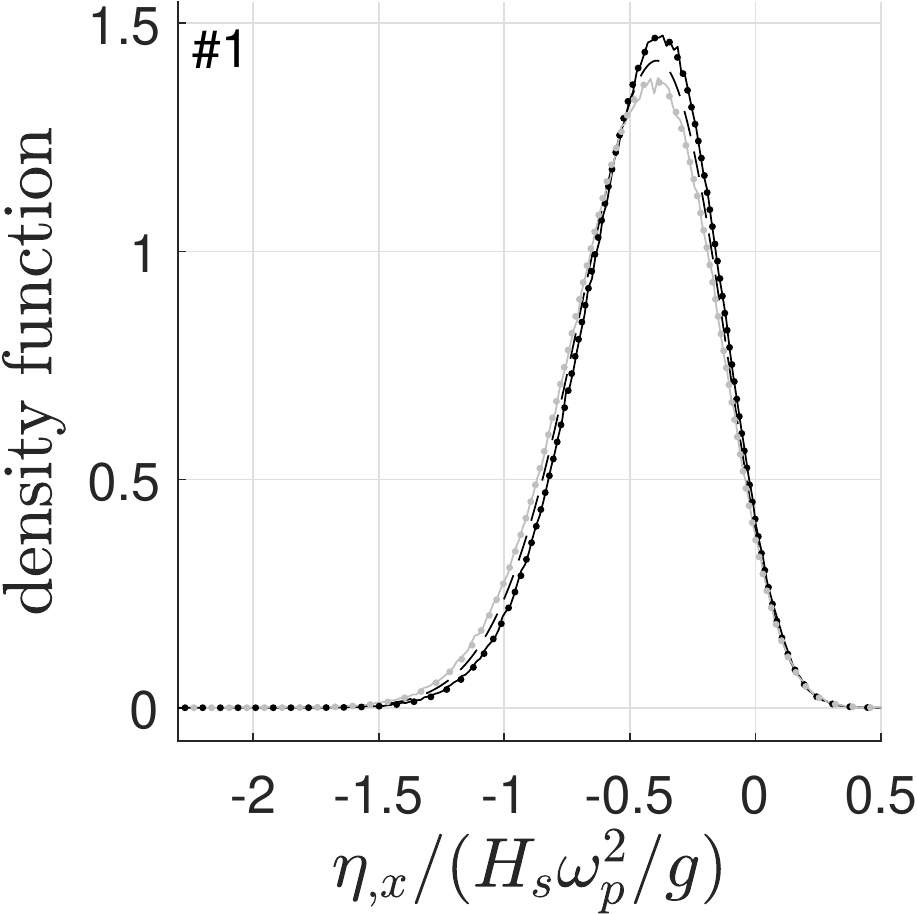} 
        &        
        \\
        \includegraphics[width=\scaleF\textwidth]{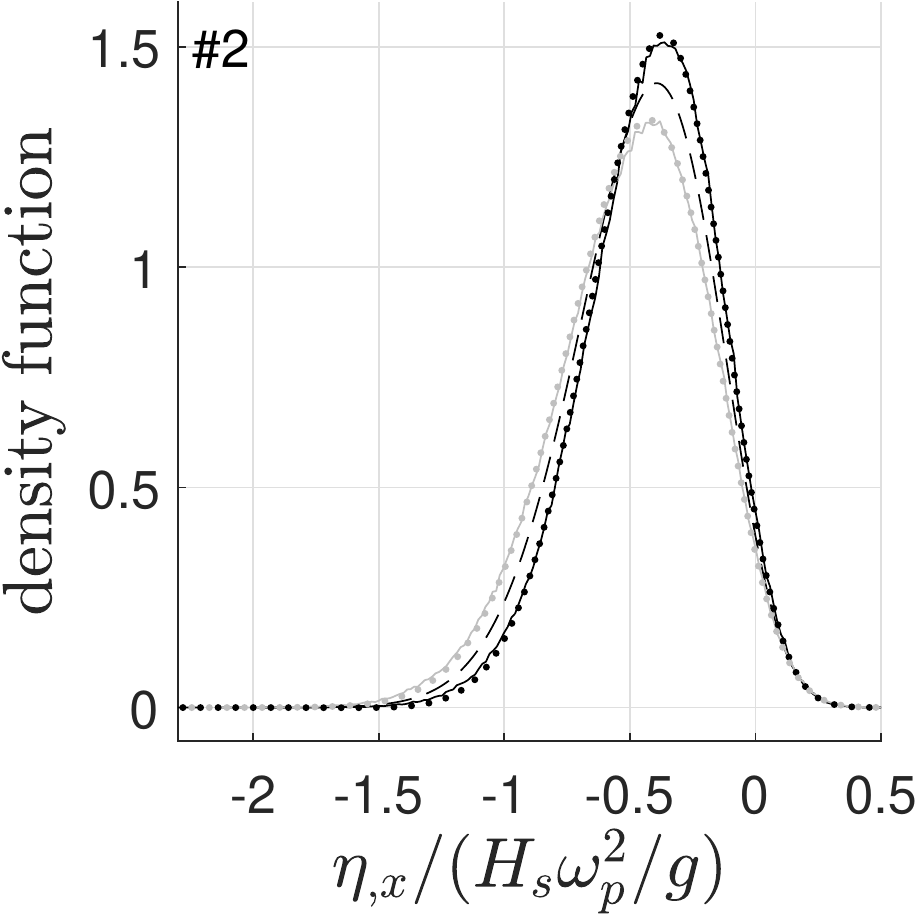} 
        & 
        \includegraphics[width=\scaleF\textwidth]{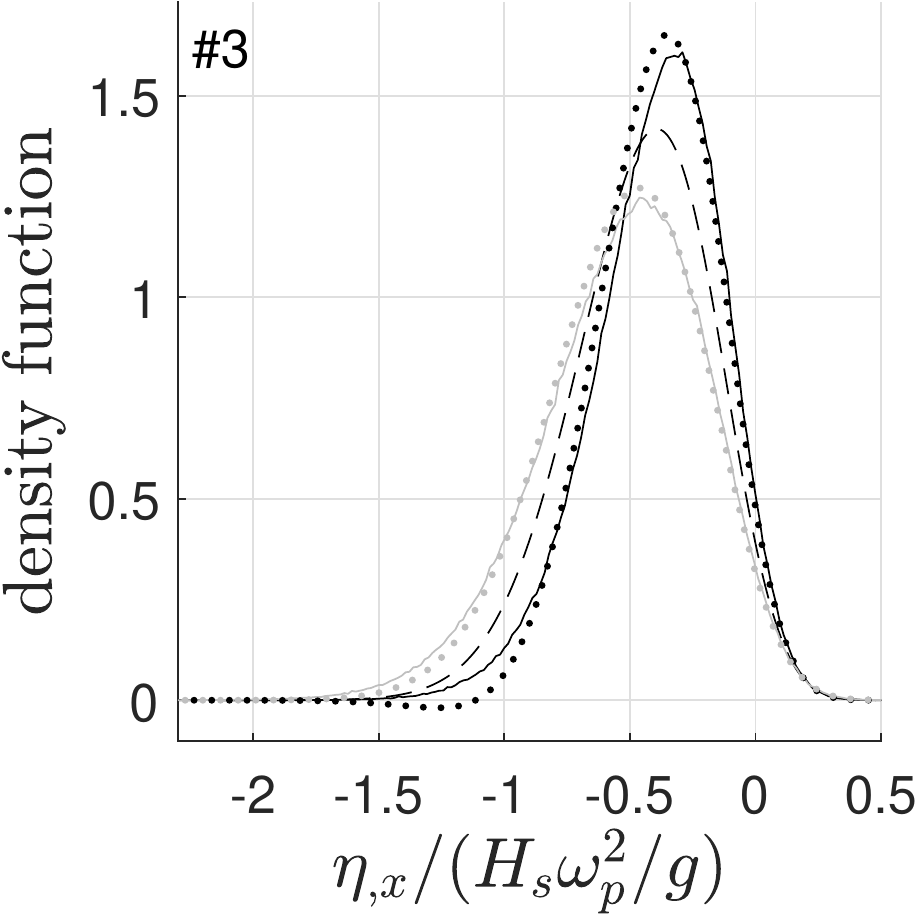} 
        &
        \includegraphics[width=\scaleF\textwidth]{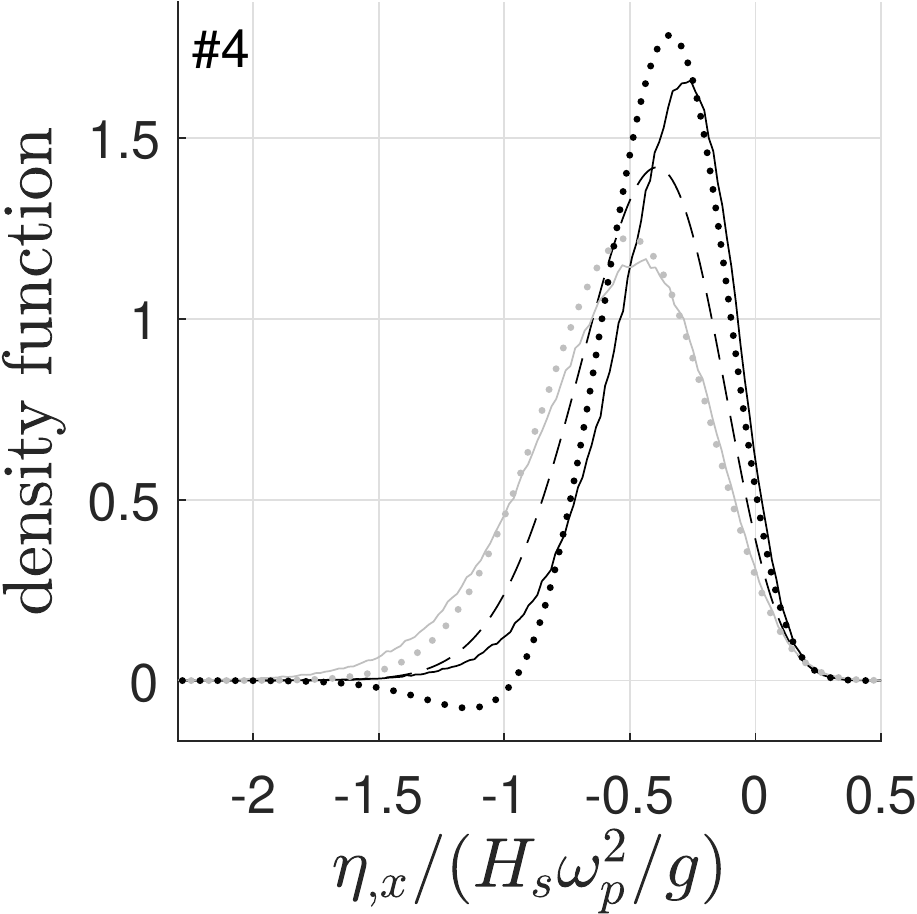} 
        \\
        \includegraphics[width=\scaleF\textwidth]{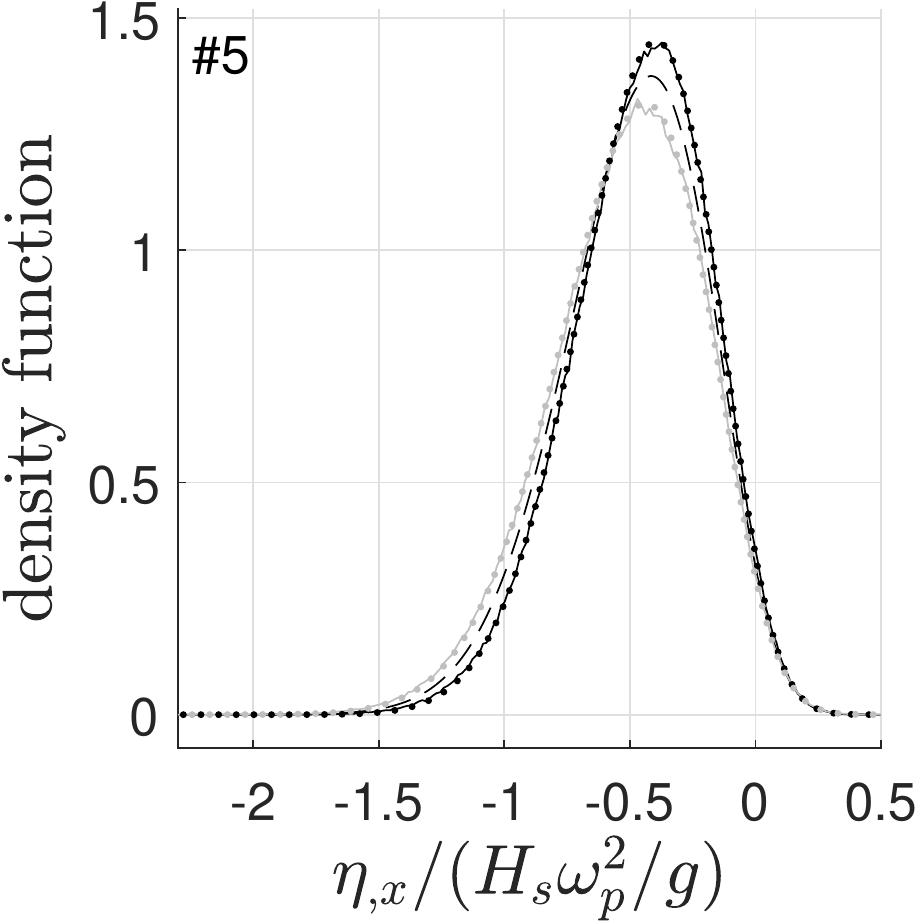} 
        & 
        \includegraphics[width=\scaleF\textwidth]{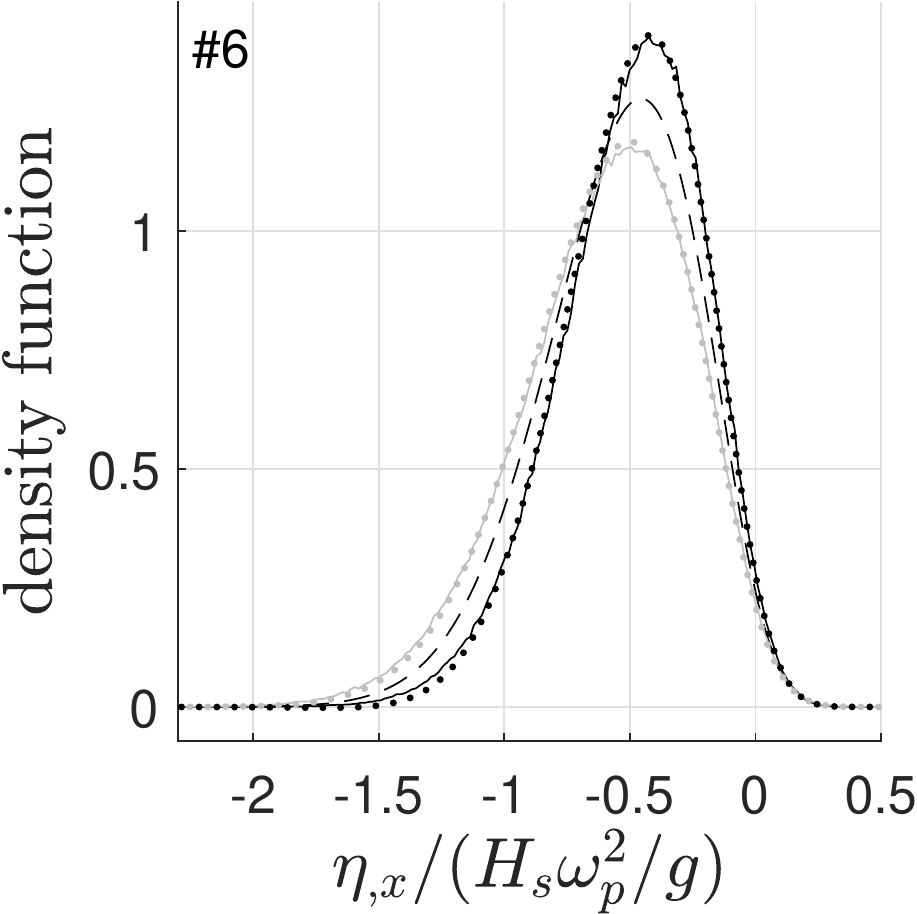} 
        &
        \includegraphics[width=\scaleF\textwidth]{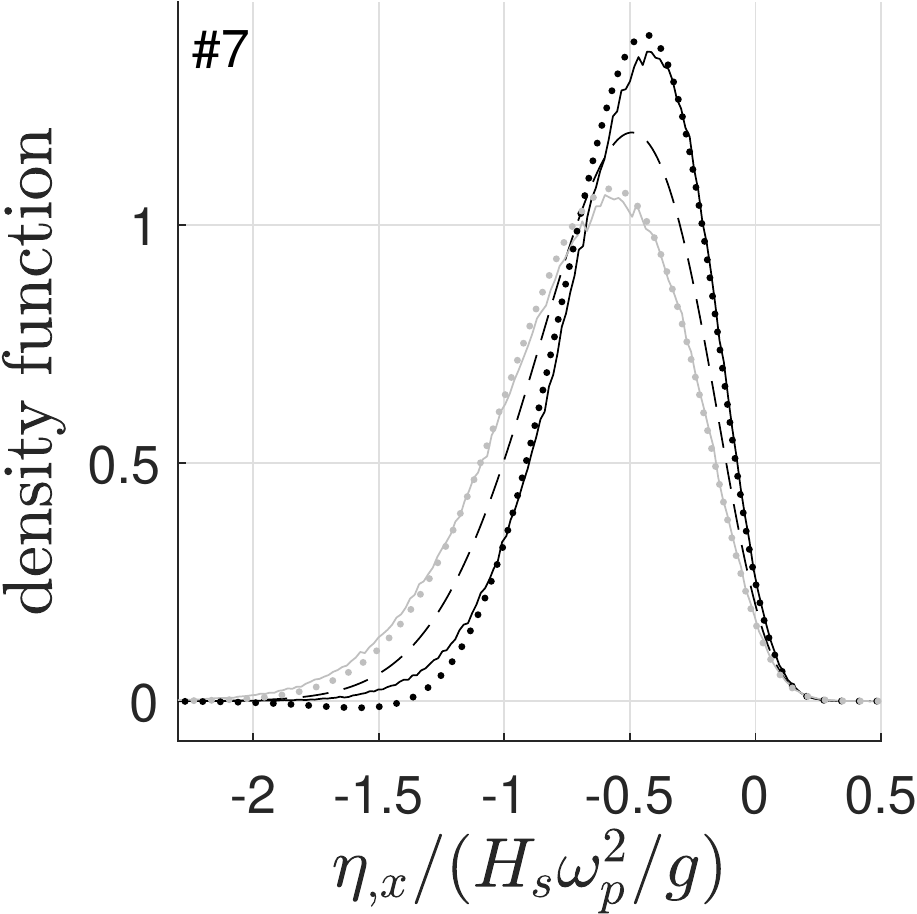} 
        \\        
\end{tabular}        
\end{center}
\caption{
Density function
of $\sx$, given  free-surface upcrossing.
    The way to read this figure is similar to the description given in the caption of \fig{fig_uz_dists}.}
\label{fig_sx_dists}
\end{figure}



\def\scaleF{0.33}
\def\scaleL{0.2}
\def\scaleSpeLine{0.75}
\def\widthSpeLine{0.1}

\begin{figure}[h!]
\begin{center}
\begin{tabular}{c} 
        \includegraphics[width=\scaleF\textwidth]{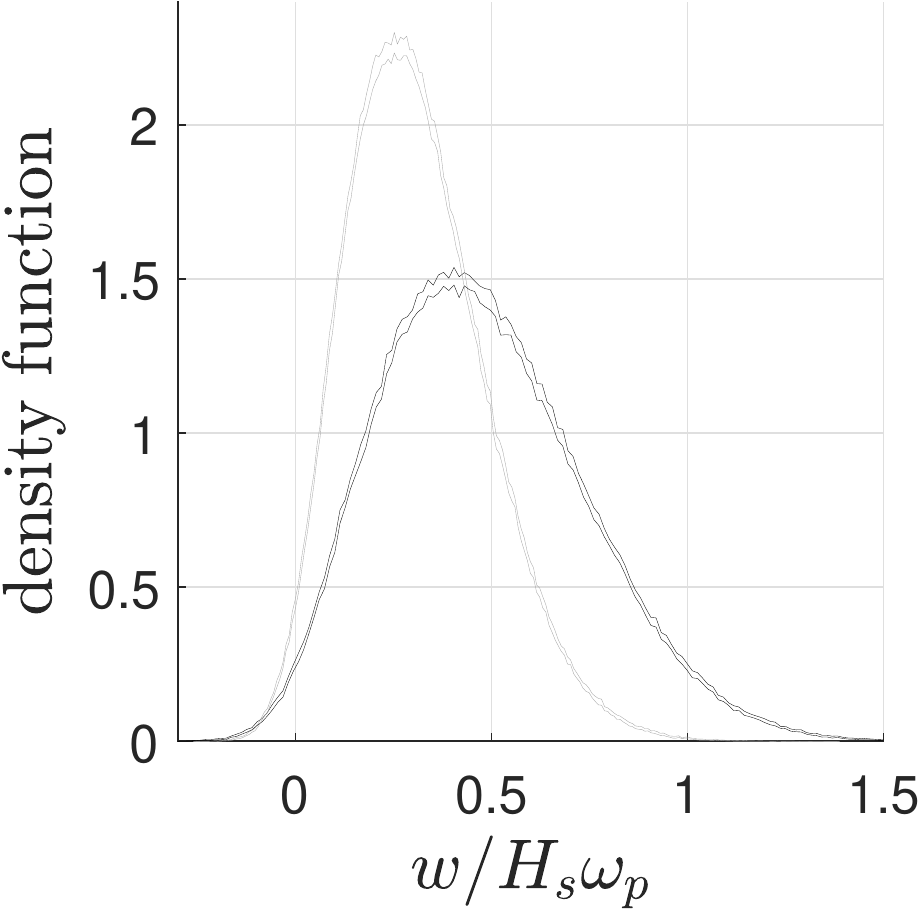} 
\end{tabular}        
\end{center}
\caption{
Statistical precision of numerical density functions. 
This figure shows an example of the two-standard-deviation envelopes ($\simeq 95\%$ confidence level when errors are Gaussian)
obtained for the Monte Carlo estimates of the density functions, given upcrossing.
Results are shown for the variable $\vz$, in the sea state configuration $\#4$,
at two different crossing level elevations: $\lc / H_s = -1/2$ (black)$;1/2$ (grey).
}
\label{fig_error_bars}
\end{figure}


Figs. \ref{fig_uz_dists}-\ref{fig_ux_dists}-\ref{fig_sx_dists} show 
the conditional probability density function, given upcrossing,
for the variables $\vz$, $\vx$, $\sx$ respectively.
In these figures, each panel corresponds to a configuration as listed in Table \ref{table_cases}.
Results within the second-order wave model are shown for two different crossing levels: 
$\lc / H_s = -1/2; 1/2$.
For these two crossing levels the analytical approximation (dotted lines) 
is compared with predictions obtained from Monte Carlo simulations (solid lines).
The prediction from the linear wave model is also shown for comparison (dashed line).
Within the linear wave model, 
the conditional distributions of $\vzL$ and $\sxL$, given upcrossing,
do not depend on the altitude of the crossing level (see \S\ref{subsubsec_considered_vars}).
Conversely, 
the conditional distribution of $\vxL$, given upcrossing, depends on the level altitude, but only through its mean.
In order to make all distributions under study independent of the crossing level within the linear wave model, 
$\vx - E[\vx^{(1)}]$, 
instead of $\vx$, is considered in Fig. \ref{fig_ux_dists}. 
Here, $E[\vx^{(1)}]$ 
should be understood as the conditional mean of $\vx^{(1)}$,
given that $\eL(t)$ upcrosses the level $\lc$.
Hence, for the three variables, Figs. \ref{fig_uz_dists}-\ref{fig_ux_dists}-\ref{fig_sx_dists} 
focus on the first-order/second-order differences.

For the sea state with the smallest magnitude of wave nonlinearities (configuration $\#1$),
the analytical approximations are in very good agreement with the Monte Carlo empirical distributions,
this for the three kinematic variables and the two crossing levels considered.
This good agreement, along with the smallness of the short fluctuations (due to sampling) seen in the Monte Calo curves,
show that the statistical precision of the Monte Carlo results is sufficient to effectively gauge the accuracy of the analytical model.
\fig{fig_error_bars} gives an alternative impression of the statistical precision achieved in Monte Carlo results, by showing 
an example of uncertainty envelopes, computed for the the Monte Carlo curves shown in \fig{fig_uz_dists}, panel \# 4.
For the sake of clarity, Monte Carlo uncertainty envelopes are not shown in Figs. \ref{fig_uz_dists}-\ref{fig_ux_dists}-\ref{fig_sx_dists}.

As the magnitude of wave nonlinearities increases 
(either due to increasing significant wave height or decreasing water depth)
the analytical approximations follow the trends of the Monte Carlo distributions 
(the latter being considered as the reference).
However, deviations between analytical and Monte Carlo distributions become 
also more and more notable as wave nonlinearities grow.
For the three variables considered, 
a range of negative densities develop in the analytical approximations,
as the magnitude of nonlinearities increases
(see panels $\#4$ in Fig. \ref{fig_uz_dists}, 
$\#2-3-4$ in Fig. \ref{fig_ux_dists}, 
$\#3-4-7$ in Fig. \ref{fig_sx_dists}). 

Without doing formal statistical tests at this stage,
the most ``robust'' analytical approximation seems to be the one obtained for $\vz$.
It follows quite faithfully the evolution of Monte Carlo distributions as wave nonlinearities grow 
(even for the most severe configurations, $\#4-7$).
Besides, the magnitude of 
spurious negative densities remains limited in the case of $\vz$.
The most ``problematic'' behavior of the analytical approximations may be the one observed for $\vx$:
as the significant wave height is increased ($\# 1-2-3-4$), for the crossing level $\lc/H_s = -1/2$,
the analytical approximation suggests that the conditional distribution qualitatively changes,
becoming bimodal.
If there were not the benchmark provided by the Monte Carlo results,
the falseness of this behavior would not be easy to identify.

The appearance of negative probability densities and spurious oscillatory features
reflects the fact that the Edgeworth approximation takes the form of a polynomial correction 
applied to a Gaussian baseline.
This issue gets amplified as the magnitude of non-Gaussianities 
-- which are the consequence of wave nonlinearities in the present context -- 
 grow in the target distribution (see e.g. \cite{ochi_1984, blinnikov_1998} for a detailed discussion about this matter).
For the sake of clarity, results in Figs. \ref{fig_uz_dists}-\ref{fig_ux_dists}-\ref{fig_sx_dists} 
are shown for only two different crossing altitudes, $\lc/H_s = -1/2;1/2$.
The examination of the analytical approximations, computed for other crossing altitudes, 
has shown that the tendency to develop spurious oscillatory features and negative densities
is amplified (resp. reduced) as the considered crossing altitude is moved away from 
(resp. brought closer to) the mean water level, $\lc = 0$.

The present
results suggest that the Edgeworth-type analytical approximation is able to capture only the main features
of the conditional distributions, given upcrossing.
This is not surprising since the analytical approximation is based on Rice's integration
of the leading-order Edgeworth expansion of the trivariate distribution of $\eta$, $\de$ and $\xi$ 
($\xi$ standing for the level-crossing conditioned variable; 
i.e. $\vz$, $\vx$, or $\sx$ in the present section).
The leading-order Edgeworth expansion takes into account cumulants only up to the third order (see Eq.~\ref{eq_EW3D}).
Then, it would be quite incidental that the resulting analytical approximation 
renders a set of probabilistic features beyond those tightly connected 
to the non-conditioned second-order and third-order cumulants of the 
triad $(\eta, \de, \xi)$.
Then, one could expect the analytical approximation to render the effect of wave nonlinearities 
on the mean, variance, and skewness of the conditional distributions, given upcrossing.
In fact, this induction is not so obvious because Rice's integration (see Eq. \ref{eq_rice_dist_gene}) 
constitutes a non-trivial transformation of the original non-conditioned trivariate distribution:
it cuts the non-conditioned distribution by the plane $\eta = \lc$ 
and then integrated along the positive crossing velocities, $\de$,
with an extra weighting prefactor equal to $\de$.
This matter is investigated in the next paragraph, 
where the conditional mean, variance and skewness, given upcrossing, 
obtained from the analytical approximation are compared 
with those estimated from the Monte Carlo experiments.

\subsubsection{Mean, variance and skewness of the conditional distributions}
\label{subsubsec_mvs_condi_dists}


\begin{sidewaysfigure}
    \centering
    \includegraphics[width=\textwidth]{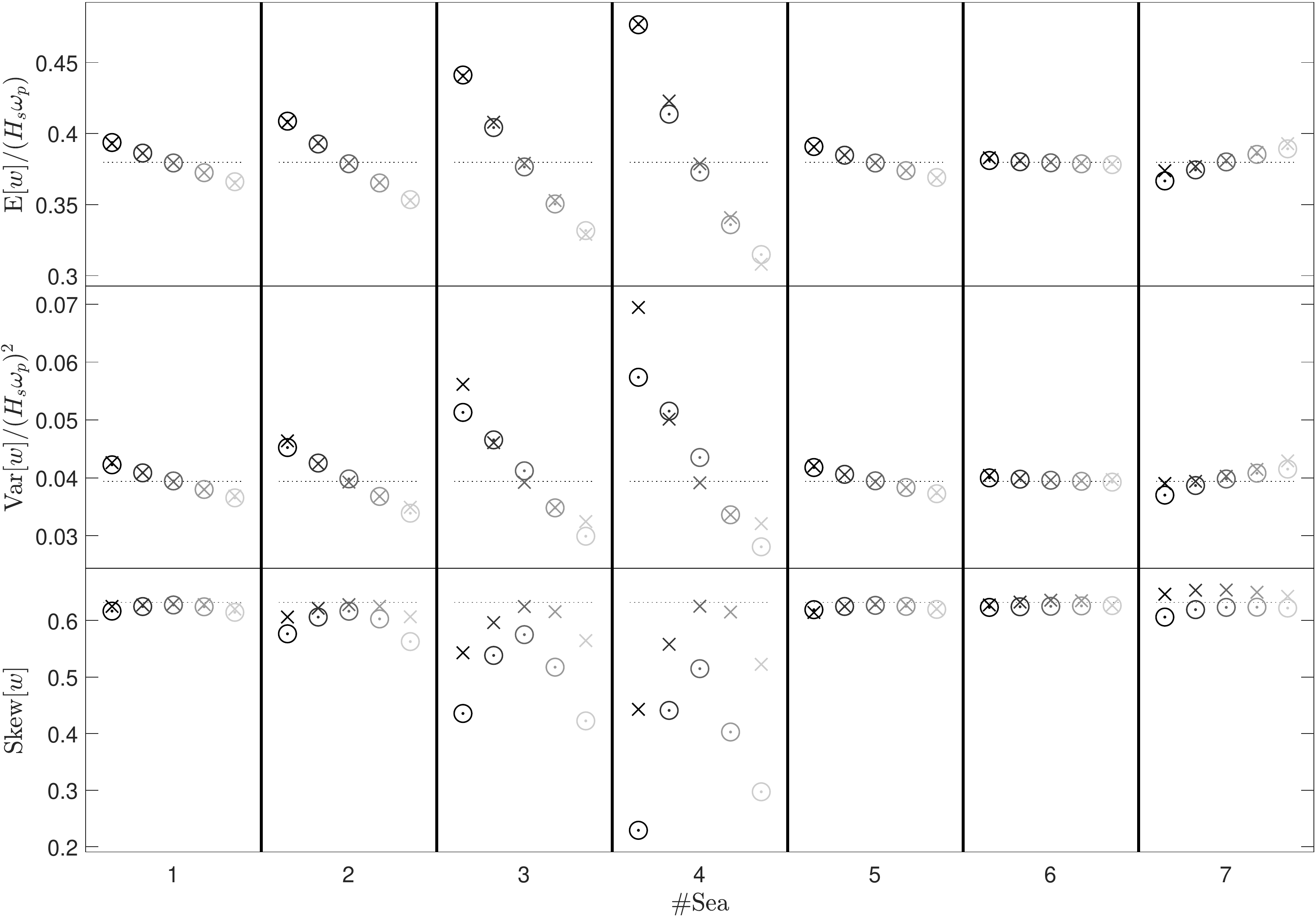}
    \caption{
    Mean, variance, and skewness of the conditional distribution of $\vz$, given upcrossing. 
    Each column, labelled with an integer at the bottom, corresponds to a sea state configuration, 
    as listed in Table \ref{table_cases}.
    The mean and the variance are given in nondimensional form.
    Different models are compared: 
    (i) linear wave model (horizontal dotted lines)
    (ii) second-order wave model, Monte Carlo simulations (crosses)
    (iii) second-order wave model, Edgeworth approximation (circled dots).
    In each panel, results are shown for different crossing levels, using a gradation of lightening gray, from the left to right: 
    $\lc / H_s = -1/2 \ {\rm (black)};-1/4;0;1/4;1/2 \ {\rm (light \ gray)}$.
    In the linear wave model, the conditional distribution of $\vz$ does not depend on the value of the crossing level,
    hence a horizontal dotted line is used.
    }
    \label{fig_uz_particulars}
\end{sidewaysfigure}



\begin{sidewaysfigure}
    \centering
    \includegraphics[width=\textwidth]{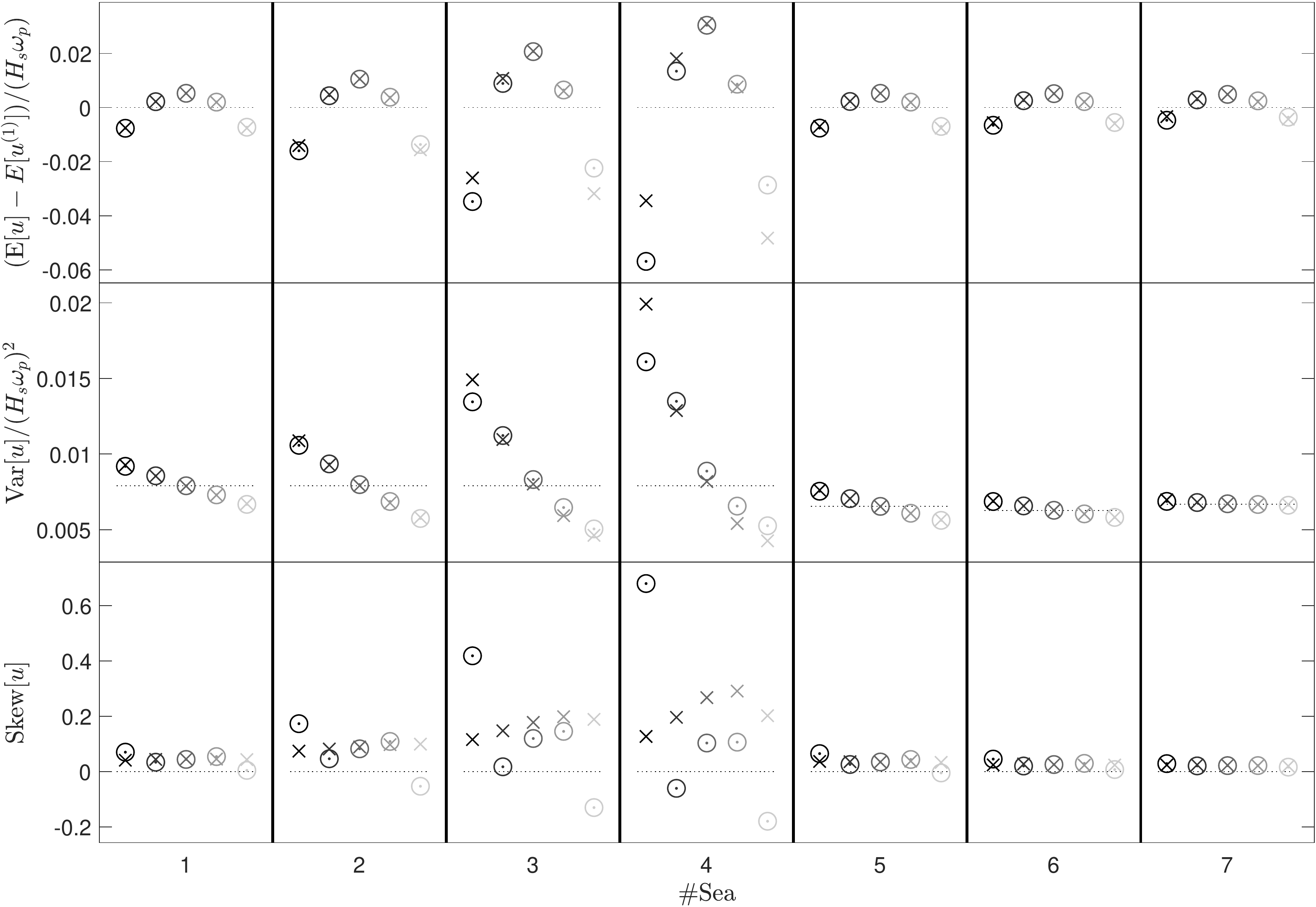}
    \caption{
    Mean, variance, and skewness of the conditional distribution of $\vx$, given upcrossing. 
    The way to read this figure is similar to the description given in the caption of \fig{fig_uz_particulars}.
    The only difference concerns the conditional mean from which the value predicted by the linear wave model, $E\left[ \vxL\right]$, has been subtracted.
    }
    \label{fig_ux_particulars}
\end{sidewaysfigure}



\begin{sidewaysfigure}
    \centering
    \includegraphics[width=\textwidth]{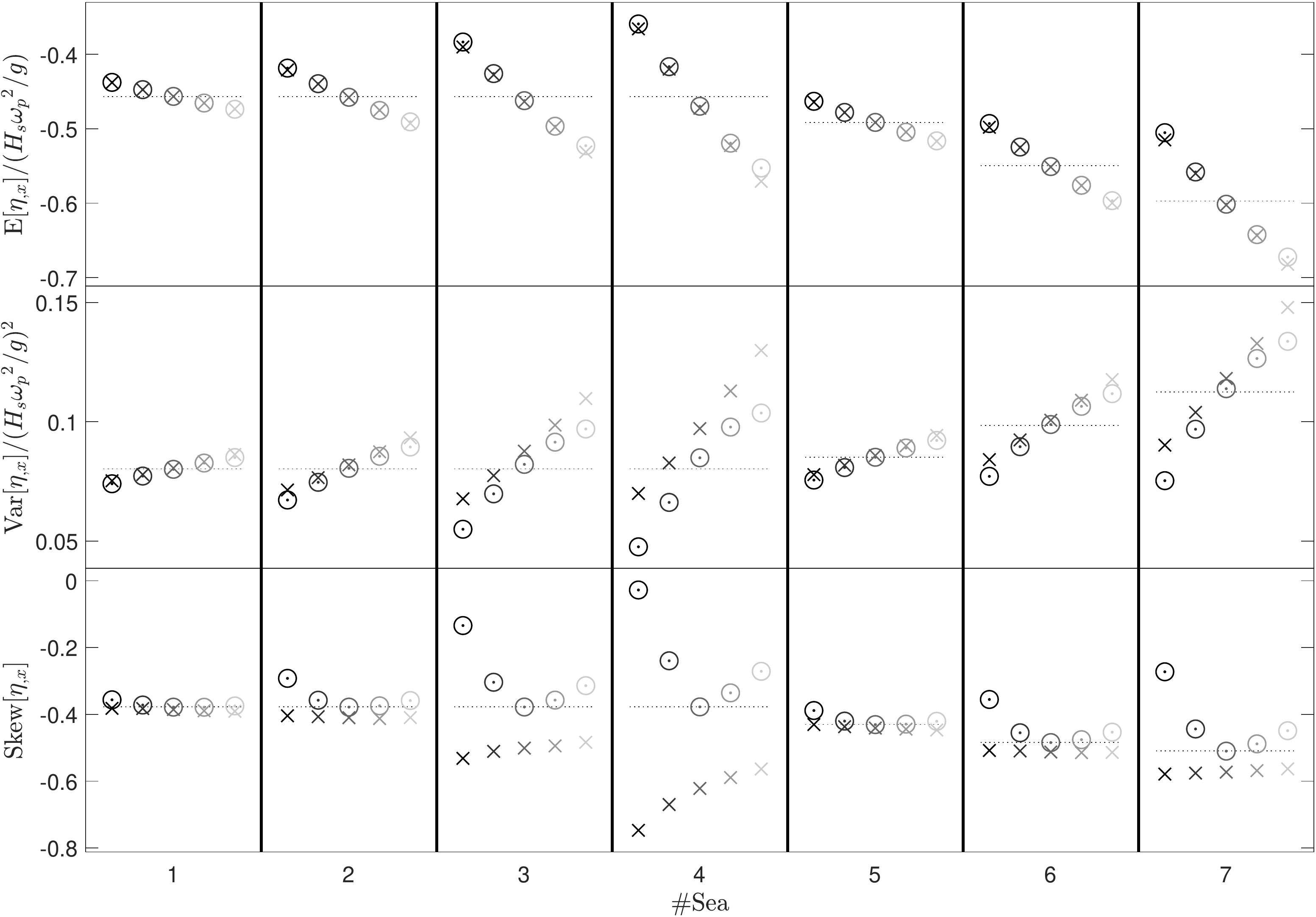}
    \caption{
    Mean, variance, and skewness of the conditional distribution of $\sx$, given upcrossing. 
    The way to read this figure is similar to the description given in the caption of \fig{fig_uz_particulars}.
    }
    \label{fig_sx_particulars}
\end{sidewaysfigure}


As  the mean, variance and skewness of the analytical approximations,
do not lend themselves to closed-form calculations,
their evaluation is performed through numerical integration.
Beforehand, a decision has to be made regarding 
the spurious negative probability densities which may develop 
in some cases.
In the present study negative densities are simply ceiled to zero.
Then the resulting function is renormalized to ensure that its integration 
over the entire sample space is equal to 1. 
This way, the mean, variance and skewness are computed for a function 
which actually meets the mathematical properties of a probability density function.

The resulting values are shown in 
Figs.~\ref{fig_uz_particulars}-\ref{fig_ux_particulars}-\ref{fig_sx_particulars}
for the variables $\vz$, $\vx$ and $\sx$ respectively. 
In each figure, the seven panel columns correspond to the configurations listed in Table \ref{table_cases},
and the three panel lines correspond to the considered statistical quantity (mean, variance, or skewness).
In a given panel, 5 different crossing levels are considered ($\lc / H_s = -1/2;-1/4;0;1/4;1/2$),
represented in increasing order, from left to right, with a lightening shading of grey.
Results obtained from the Edgeworth-type approximations (resp. Monte Carlo experiments)
are shown as circled dots (resp. crosses).
Similarly to \S\ref{subsubsec_prob_densi}, the mean of the variable $\vx - E[\vxL]$
is considered instead of the mean of $\vx$,
(recall that $E[\vxL]$ 
denotes the conditional mean of $\vxL$, given that $\eL(t)$ upcrosses the level $\lc$).
For comparison, the values predicted by the linear wave model 
are shown as horizontal dotted lines
(since they do not depend on the actual crossing level, $\lc$).

For the conditional mean, 
the analytical approximation is overall in good agreement with the Monte Carlo estimates,
even for the most severe configurations in terms of wave nonlinearities.
The largest discrepancy is seen for the variable $\vx$, configuration $\# 4$, 
at the crossing levels $\lc / H_s = -1/2;+1/2$.
This is due to the oscillations and negative densities 
appearing in the analytical approximation of the probability density function 
(see \fig{fig_ux_dists}, panel $\#4$).

For the conditional variance,
the values predicted by the analytical approximation are also in good agreement
with the Monte Carlo estimates, for most considered cases.
Still, for some cases, the discrepancies between the analytical approximation and the Monte Carlo results
are more prominent than for the conditional mean, especially for the sea state with the most severe wave steepness 
(configuration $\#4$).
Similarly to the conditional mean, the appearance of these discrepancies are related (at least partly)
to the development of oscillations and negative densities in  
the analytical approximation
of the probability density function.
Albeit these discrepancies in some cases, 
the analytical approximation still provide the correct evolution trends 
as a function of the crossing level,
even for the most severe sea states in terms of nonlinearities.
 
Finally, regarding the 
conditional
skewness, 
the results provided by the analytical approximation are less satisfactory than for 
the conditional mean and variance.
For the kinematic variable $\vz$,
the analytical approximation still provides the correct trend as a function of the crossing level,
but with an offset increasing as wave nonlinearities grow 
(either due to an increasing significant wave height or a decreasing water depth).
When considering the kinematic variable $\vx$, Monte Carlo results show 
that second-order wave nonlinearities have a moderate (for nonlinearities due to an increasing significant wave height)
to minor (nonlinearities due to a decreasing water depth) effect on the conditional skewness.
This trend is reproduced by the analytical approximation for the varying water depth series (configurations $\# 1-5-6-7$)
but not for the varying $H_s$ series (configurations $\# 1-2-3-4$),
where an increasing data dispersion can be seen in the analytical results.
This dispersion is the signature from the oscillations and negative densities appearing in the 
analytical approximation of the probability density function (see \fig{fig_ux_dists}).
The impact is large on the skewness, since this statistical quantity is a measure of the distribution asymmetry.
The situation is even less satisfactory for the skewness of $\sx$.
For this variable, the analytical approximation tends to predict corrections, relative to the linear model, 
which are of opposite signs compared to Monte Carlo results.

\section{Discussion}
\label{sec_discussion}

An analytical approximation for the conditional distribution of a wave kinematic variable, $\xi$, given free-surface upcrossing,
has been derived for the second-order wave model. 
The analytical approximation is based on the application of Rice's formula 
to the Edgeworth expansion 
of the trivariate distribution of $(\eta, \de, \xi)$, truncated to the leading-order.
The scope of the analytical model can be readily extended regarding three aspects.
(i) When the interest would be in downcrossing events only, or both upcrossing and downcrossing events,
the analytical formulae derived in the present study may be readily adapted.
(ii) 
Upcrossing events could be monitored for a second-order process 
which is different from the free-surface elevation measured at a fixed station, $\eta(t)$.
The analytical model would remain readily applicable,
as long as the linear and quadratic transfer functions of the considered process are tractable.
For instance free-surface crossing events could be monitored for a material point with forward motion 
(see \cite{hascoet_2021} for a recent investigation of the subject within the linear wave model) 
and/or seakeeping motions.
Free-surface crossing events could also be checked in space, at a fixed time, along a given direction.
Alternatively, 
crossing events could be monitored for a quantity different from the free-surface elevation
(e.g. the stress response of a marine structure).
(iii) The application of the Edgeworth-based approximations derived in Sections \ref{sect_rice_ew}
is not restricted to second-order processes. 
They may be applied to any couple of interdependent stochastic processes (one being conditioned to the level-crossing of the other),
as long as the necessary third-order cumulants can be estimated.

\subsection{On the experimental context}

It is worth noting that the non-conditional joint cumulants, $K_{abc}$,
may be directly measured from experimental data (in the field or in a laboratory), 
without relying on a specific wave theory.
This would necessitate simultaneous observations of the free-surface elevation, $\eta$, the elevation rate $\dot{\eta}$, 
and a third wave variable $\xi$ (for which the conditional distribution, given upcrossing, is wanted).
``Non-conditional'' in the experimental context, 
means that observations should be done on a grid in space and/or time, which is fixed \textit{a priori} -- 
such observations may be said asynchronous.
Then, from these empirical estimates of non-conditional joint cumulants, 
the present analytical model may be used to compute an approximation for the conditional distribution of $\xi$, 
given free-surface upcrossing.
Such an approach may prove useful, 
as synchronous measurements of $\xi$ 
-- measurements coinciding with upcrossing events -- 
may be more challenging to achieve, depending on the experimental context.
Alternatively, instead of directly estimating the cumulants $K_{abc}$, 
the QTFs of $\eta$, $\dot{\eta}$, $\xi$, could be empirically estimated, by using a bispectrum analysis;
such an approach was applied by Ochi and Ahn \cite{ochi_1994} to a free-surface elevation record.
Then the cumulants may be estimated from these empirical QTFs by using Eqs. (\ref{eq_cumulants_O2}-\ref{eq_cumulants_O3}).

Experimental measurements may be also useful to further test the analytical model derived in the present paper.
To the best knowledge of the author, no experimental study has estimated
non-conditional trivariate joint cumulants in a wave field,
or has directly estimated the conditional distribution of a wave variable, 
given free-surface upcrossing.
Several studies have experimentally investigated 
the joint distribution of sea slopes, in offshore experiments (e.g. \cite{cox_1956, longuet_1963} using sun glitter photographs, \cite{mironov_2012} using stereo images) or in tank test experiments (e.g. \cite{zavadsky_2017} using a slope gauge and stereo imaging), as it may be of practical importance for remote sensing applications.
Huang et al. (1984) \cite{huang_1984} have experimentally 
investigated the joint distribution of slope and elevation in model experiments, 
by using an assembly of wave probes.
Several studies have also experimentally investigated fluid kinematics in irregular waves, 
both at model scale and full scale (see e.g. \cite{gudmestad_1993} and references therein).
Most of these studies focus on the horizontal fluid velocity, 
as it is a key quantity to accurately estimate the wave-induced drag on offshore structures such as compliant platforms.
Many of these studies also focus on the kinematics in high and steep waves (e.g. \cite{grue_2003} using particle image velocimetry in model-scale experiments,
\cite{choi_2007} using  particle image velocimetry and laser Doppler velocimetry in model-scale experiments), 
bearing in mind the safety of marine structures with respect to extreme waves.
The investigation of the stochastic properties of wave kinematics, in a given sea state, 
is scarcer in the literature. 
For instance, Skjelbreia et al. (1989, 1991) \cite{skjelbreia_1989, skjelbreia_1991} used laser Doppler velocimetry 
to investigate the statistics of wave kinematics at different altitudes above and below the mean water level. 
Measuring wave kinematics at a fixed point near the mean water level
usually excludes the use of current-meters, 
as their accuracy is degraded by frequent emergence  and reimmersion events.
In such a configuration, the use of optical techniques such as laser Doppler velocimetry or particle image velocimetry in test tank experiments
is preferable. 
Intermittent emergence will also bring complications in the statistical analysis of data,
as the time series will be ``undefined'' during the periods where
the measurement point lies outside the water domain (i.e. lies above the free surface).
Cie{\'s}likiewicz \& Gudmestad (1993) \cite{cieslikiewicz_1993} 
have investigated this problem in the ``forward'' way 
(i.e. computing submergence-biased distributions from unbiased distributions) 
within the second-order wave model.
One solution to avoid complications due to intermittent emergence, 
is to measure the fluid velocity at a point which keeps its instantaneous immersion depth constant,
by following the vertical motion of the free surface. 
Such measurements may be achieved thanks to particle image velocimetry 
or a current meter mounted on a movable frame (see e.g. \cite{klinting_1990, donelan_1992}).
Whether in the field or in a laboratory, 
one of the biggest challenges may be to acquire a sufficiently large amount of data,
to achieve a satisfactory statistical precision in the estimates of joint cumulants and/or distributions.

It is important to note that comparisons of experimental data with the model derived in the present study
may be carried out at two different levels, separately.
At a first level, 
the comparisons may focus on the ability of the second-order wave model to render the stochastic properties of irregular seas.
In this case, empirical estimates of the joint cumulants 
may be directly compared with the predictions from the second-order wave model,
without necessarily considering the question of crossing conditioning.
At a second level, 
a comparison with experimental results may focus on the ability of the Edgeworth-based approximation
to infer the effect of crossing conditioning from non-conditional measurements.
For this purpose, the non-conditional joint cumulants, $K_{abc}$, may be empirically estimated,
and used to compute the Edgeworth-based approximation of the conditional distribution of a wave variable $\xi$, 
given free-surface crossing.
Then, if the conditional distribution, given upcrossing, can be also directly estimated from synchronous measurements,
the results from both approaches may be compared.

\subsection{On the use of the stochastic model in the context of slamming and green water}

One of the main applications of
level-crossing conditioning, in the marine context, 
may be the investigation of stochastic slamming or green water, on ships or marine structures.
Both the frequency of these events and the conditional distribution of the incoming wave kinematics, 
given such an event occurs, are of practical interest.
This matter can be addressed by modeling free-surface upcrossings
at a ``control'' point attached to the considered marine platform 
-- for instance the lowest point of a marine substructure exposed to wave slamming, 
or a critical flooding point on a ship's deck exposed to green water. 

Such an approach has been implemented to investigate stochastic slamming 
within the framework of the linear wave model (see e.g. \cite{ochi_1971, helmers_2012, hascoet_2020}).
The analytical model derived in the present study may be used 
to include the effect of second-order wave nonlinearities in such analyses.
The seakeeping motions of a marine platform and its diffracted waves, 
may be also included in the analysis, 
as long as the related transfer functions and quadratic transfer functions
can be estimated (experimentally or numerically).
Note however that the present analytical approximation 
only renders the ``core'' of the distributions, 
and becomes inaccurate toward the tails of distributions. 
This means that the Edgeworth-based approach would not be suitable to
reckon the risk of failure due to extreme slamming events -- 
i.e. events which would lead to failure through the overshoot of the yield or ultimate strength of the structure material.
Conversely, the model may be appropriate to estimate the risk of failure due to fatigue,
as the cumulated fatigue damage may be dominated by a large number of repeated slamming events.

The vertical component of the fluid velocity, at impact, 
is usually considered as the most decisive variable 
when estimating the resulting slamming loads (see e.g. \cite{ochi_1973, rassinot_1995, wang_2002, hermundstad_2007, wang_2016}).
Depending on the shape of the slamming-exposed body, 
it may be important to include additional kinematic variables 
(for instance the free-surface slope, the horizontal fluid velocity, or the fluid acceleration) 
as input of the considered slamming model (see e.g. \cite{helmers_2012, hascoet_2020}).
The question then arises as to whether the model derived in the present paper may be extended to 
approximate the joint distribution of multiple variables, given upcrossing.
In the particular case where the interest would be in the joint distribution of $(\de,\xi)$, given upcrossing, 
its Edgeworth approximation is readily obtained from
\begin{equation}
\label{eq_ew_condDist_2D}
\hat{f}_{\de,\xi | \eta \uparrow \lc} (\de, \xi) 
= \frac{1}{\st {\sigma_{\de}}^2} \frac{ \displaystyle  \de  \hat{f}_{\xv\yv\zv}(\tilde{\lc}, \de / \sd ,\xi / \st)   }{
\displaystyle \int_0^{+\infty} \yvi \hat{f}_{\xv,\yv} (\tilde{\lc},\yvi) \ {\rm d} \yvi } \, , \ \de>0 \, .
\end{equation}
All the material necessary to explicitly compute \eq{eq_ew_condDist_2D} 
is provided in the present study,
and no further development is required.
For other cases -- i.e. conditional bivariate distributions with $\de$ not being one of the considered random variables, 
or conditional joint distributions of dimension larger than two -- 
the non-conditional distribution to be considered (before applying Rice's formula), 
would be of dimension larger than three.
To the leading order, the Edgeworth correction to a \mbox{$N$-dimensional} Gaussian distribution,
with $N>3$,
would still be an expression involving third-order cumulants and third-order Hermite polynomials,
similar to the expression appearing into square brackets in \eq{eq_EW3D}. 
Then, Rice's integration
may be carried out by using a method similar to the one presented 
in Appendices B-C-D.
Note however that the Hermite polynomials would have a different expression 
since the baseline distribution from which they are defined (see Eq. \ref{eq_hermite_gene}) 
would no longer be a trivariate Gaussian distribution,
but its N-dimensional analog.
For instance, if a fourth variable, $\upsilon$, were to be introduced in the problem, 
one would obtain $H_{abc0}(\xv,\yv,\zv, \upsilon) \ne H_{abc}(\xv,\yv,\zv)$.
As the considered dimension $N$ increases, 
the analytical developments are expected to inflate quickly, 
and a computer algebra system may be of help.


\setcounter{section}{0}
\renewcommand\thesection{Appendix \Alph{section}}
\renewcommand\thesubsection{\Alph{section}.\arabic{subsection}}

\section{Hermite polynomials}
\label{sec_ap_hermite}

\subsection{Univariate polynomials}
\label{subsec_hermite_1D}

The probabilist's Hermite polynomials are defined as follows:
\begin{equation}
H_a (x) = (-1)^a \exp{x^2/2} \frac{{\rm d}^a}{{\rm d} x^a} \exp{-x^2/2} \, .
\end{equation}
The first polynomials, up to the third order, are given by
\begin{align}
H_0(\chi) & =  1 \, , \\
H_1(\chi) & =  \chi \, , \\
H_2(\chi) & =  \chi^2-1 \, , \\
H_3(\chi) & =  \chi^3-3\chi \, . \\
\end{align}

\subsection{Trivariate Hermite polynomials}
\label{subsec_tri_hermite}

The trivariate probabilist's Hermite polynomials are defined by,
\begin{equation}
\label{eq_hermite_3D_general}
H_{abc} (\xvg,\yvg,\zvg) = \frac{(-1)^{a+b+c}}{f_{X}(\xvg,\yvg,\zvg)} 
\frac{\partial^a}{\partial {\xvg}^a} \frac{\partial^b}{\partial {\yvg}^b} \frac{\partial^c}{\partial {\zvg}^c} 
f_{X} (\xvg,\yvg,\zvg) \,
\end{equation}
where $f_{X}$ is the density function 
of the trivariate standard normal distribution.
In the present study, the variables $\eta$ and $\de$ are uncorrelated, 
and  \eq{eq_hermite_3D_general} may be particularized 
into \eq{eq_hermite_gene}.
From Eqs. (\ref{eq_J3}-\ref{eq_hermite_gene}), the first-order polynomials 
can be readily computed:
\begin{align}
\label{eq_hernite_h100}
H_{100}(\xv,\yv,\zv) & =  \displaystyle \left[ (1-\rhp^2)\xv + \rh\rhp\yv - \rh \zv \right] / \delta_3 \, , \\
\label{eq_hernite_h010}
H_{010}(\xv,\yv,\zv) & =  \displaystyle \left[ \rh\rhp\xv + (1-\rh^2)\yv - \rhp \zv \right] / \delta_3 \, , \\
\label{eq_hernite_h001}              
H_{001}(\xv,\yv,\zv) & =  \displaystyle \left[ - \rh\xv - \rhp\yv + \zv \right] / \delta_3 \, ,
\end{align}
where $\delta_3$ is a numerical factor defined by
\begin{equation}
\label{eq_d3}
\delta_3 = 1-\rh^2-\rhp^2 \, .
\end{equation}
Using the recurrence relations 
\begin{align}
\displaystyle
 H_{abc} & = \displaystyle - \frac{\partial}{\partial \xv}H_{a-1bc}  + H_{a-1bc} H_{100} \, , \\
 H_{abc} & = \displaystyle - \frac{\partial}{\partial \yv}H_{ab-1c}  + H_{ab-1c} H_{010} \, , \\
 H_{abc} & = \displaystyle - \frac{\partial}{\partial \zv}H_{abc-1}  + H_{abc-1} H_{001} \, ,      
\end{align}
higher-order Hermite polynomials may be conveniently expressed in terms of first-order polynomials.
However, as calculations will show, 
an explicit expression for higher-order Hermite polynomials is not required in the present study.

\section{Computation of the integrals $\mathcal{I}_{abc}$}
\label{sec_iabc}

The level of complexity to compute the integrals $\mathcal{I}_{abc}$ (see Eq. \ref{eq_Iabc})
depends on the value of $b$.
Three different cases need to be considered: $b=0$, $b=1$, $b\ge2$.

\subsection{Computation of $\mathcal{I}_{abc}$ with $b\ge2$}
\label{subsec_Iabc_b2}

The case where $b \ge 2$ is the simplest one. An integration by parts yields
\begin{equation}
\mathcal{I}_{abc}(\tilde{\lc},\zeta) = 
- \left[ \yvi H_{ab-1c}(\tilde{\lc},\yvi,\zeta)J_{3}(\tilde{\lc},\yvi,\zeta) \right]_{\yvi=0}^{\yvi=+\infty}
+ \int_0^{+\infty} H_{ab-1c}(\tilde{\lc},\yvi,\zeta)J_{3}(\tilde{\lc},\yvi,\zeta) {\rm d} \yvi \, .
\end{equation}
The first term is equal to zero and the second term directly yields
\begin{equation}
\mathcal{I}_{abc}(\tilde{\lc},\zeta) = 
- \left[ H_{ab-2c}(\tilde{\lc},\yvi,\zeta)J_{3}(\tilde{\lc},\yvi,\zeta) \right]_{\yvi=0}^{\yvi=+\infty} \, ,
\end{equation}
leading to
\begin{equation}
\mathcal{I}_{abc}(\tilde{\lc},\zeta) = 
H_{ab-2c}(\tilde{\lc},0,\zeta)J_{3}(\tilde{\lc},0,\zeta)  \, .
\end{equation}

\subsection{Computation of $\mathcal{I}_{a1c}$}
\label{subsec_Iabc_b1}

In the case $b=1$, the same integration by parts, as for the case $b\ge2$, yields
\begin{equation}
\mathcal{I}_{abc}(\tilde{\lc},\zeta) = 
- \left[ \yvi H_{a0c}(\tilde{\lc},\yvi,\zeta)J_{3}(\tilde{\lc},\yvi,\zeta) \right]_{\yvi=0}^{\yvi=+\infty}
+ \int_0^{+\infty} H_{a0c}(\tilde{\lc},\yvi,\zeta)J_{3}(\tilde{\lc},\yvi,\zeta) {\rm d} \yvi \, .
\end{equation}
The first term is equal to zero. 
The second term may not be readily computed at first sight.
One possible way, as initially proposed by Longuet-Higgins \cite{longuet_1964} 
and further extended by Jensen \cite{jensen_1996},
is to make use of Hermite polynomial algebra, to substitute $H_{a0c}$ 
with the sum of other Hermite polynomials $H_{abc}$ with $b\ge1$,
plus an additional polynomial which does not depend on the integration variable, $\tau$.
If $b=1$, then $a+c = 2 $, 
and substitutions are needed for the polynomials $H_{200}$, $H_{002}$ and $H_{101}$.
The way forward to obtain these substitutions is developed in \ref{sec_substitutions_hermite}.
For example, considering the integral 
$\mathcal{I}_{012}$, the substitution derived for $H_{002}$ (see Eq. \ref{eq_sub_ac_2_h002}) 
can be used to obtain the following equality
\begin{align}
\mathcal{I}_{012}(\tilde{\lc},\zeta) = & \int_0^{+\infty} H_{002}(\tilde{\lc},\yvi,\zeta)J_{3}(\tilde{\lc},\yvi,\zeta) {\rm d} \yvi \nonumber \\
  =  & \int_0^{+\infty} \left\{-2 s H_{011}(\tilde{\lc},\yvi,\zeta) - s^2 H_{020}(\tilde{\lc},\yvi,\zeta) 
+ P_{02}(\tilde{\lc},\zeta)  \right\}J_{3}(\tilde{\lc},\yvi,\zeta) {\rm d} \yvi \, ,
\end{align}
which then readily yields
\begin{equation}
\mathcal{I}_{012}(\tilde{\lc},\zeta) = 
\left\{ -2s H_{001}(\tilde{\lc},0,\zeta) - s^2 H_{010}(\tilde{\lc},0,\zeta) \right\} J_{3}(\tilde{\lc},0,\zeta) 
+ P_{02}(\tilde{\lc},\zeta) G_0(\tilde{\lc},\zv) \, ,
\end{equation}
where the coefficient $s$ is defined by \eq{eq_sc} and $G_0$ is defined by \eq{eq_G0_general}.

\subsection{Computation of $\mathcal{I}_{a0c}$}

The analytical calculation of $\mathcal{I}_{a0c}$ 
may be carried out by using the same trick as for the calculation of $\mathcal{I}_{a1c}$.
But this time, a substitution should be found for the Hermite polynomials 
$H_{a0b}$ with $a+b =3$. The suitable substitutions are derived in Appendix \ref{subsec_Ia0c}.
For example, considering the integral $\mathcal{I}_{300}$, 
the substitution derived for $H_{300}$ (see Eq. \ref{eq_sub_ac_3_h300}) gives
\begin{align}
\begin{split}
\mathcal{I}_{300}(\tilde{\lc},\zeta) = \int_0^{+\infty}   \yvi \bigg\{ &
-3r H_{210}(\tilde{\lc}, \yvi,\zeta) - 3 r^2 H_{120} (\tilde{\lc}, \yvi,\zeta) \\
 & - r^3 H_{030}(\tilde{\lc}, \yvi,\zeta) + P_{30}(\tilde{\lc},\zeta)
\bigg\} J_3(\tilde{\lc}, \yvi,\zeta) \ {\rm d} \yvi \, ,
\end{split}
\end{align}
which can also be written as
\begin{align}
\mathcal{I}_{300}(\tilde{\lc},\zeta) = 
-3r \mathcal{I}_{210}(\tilde{\lc},\zeta) - 3 r^2 \mathcal{I}_{120} (\tilde{\lc},\zeta) 
 - r^3 \mathcal{I}_{030}(\tilde{\lc}, \yvi,\zeta) + P_{30}(\tilde{\lc},\zeta) G_1(\tilde{\lc},\zeta) \, .
\end{align}
The numerical coefficient $r$ is defined in \eq{eq_rc},
and the function $G_1$ is defined in \eq{eq_G1_general}.
Then, the way forward to compute the terms $\mathcal{I}_{abc}$ with $b=1$ and $b\ge2$ 
has been already presented in Appendices \ref{subsec_Iabc_b1} and \ref{subsec_Iabc_b2}, respectively.

\section{Substitutions for Hermite polynomials of the type $H_{a0c}$}
\label{sec_substitutions_hermite}

This appendix presents how Hermite polynomials $H_{a0c}(\xv,\yv,\zv)$ may be 
expressed as the sum of other Hermite polynomials $H_{abc}(\xv,\yv,\zv)$ with $b\ge 1$,
plus an additional polynomial which does not depend on $\yv$.
For this purpose, let us consider the two following differential operators:
\begin{align}
\mathcal{R} = & - \left( \frac{\partial}{\partial \xv} + r \frac{\partial}{\partial \yv} \right)  \, , \\
\mathcal{S} = & - \left( \frac{\partial}{\partial \zv} + s \frac{\partial}{\partial \yv} \right) \, ,
\end{align}
where $r$ and $s$ are constant coefficients given in Eqs. (\ref{eq_rc}-\ref{eq_sc}).
The application of these differential operators to $J_3$ gives the relations
\begin{align}
\label{eq_r_operators_apply1}
\mathcal{R}(J_3) = R \times J_3  \, , \\
\label{eq_s_operators_apply1}
\mathcal{S}(J_3) = S \times J_3 \, ,
\end{align}
where $R$ and $S$ are polynomials which do not depend on $\yv$ 
(their expression is given in Eqs. \ref{eq_r_polynomials}-\ref{eq_s_polynomials}).
On the other hand Eqs. (\ref{eq_r_operators_apply1}-\ref{eq_s_operators_apply1}) may be also expressed as
\begin{align}
\mathcal{R}(J_3) =  & \left[ H_{100} + r H_{010} \right] \times J_3 \, ,  \\
\mathcal{S}(J_3) = & \left[ H_{001} + s H_{010} \right] \times J_3 \, ,
\end{align}
which yields by identification the relations
\begin{align}
H_{100} + r H_{010} = & R \, ,  \\
H_{001} + s H_{010} = & S \, .
\end{align}
As the the operators $\mathcal{R}$ and $\mathcal{S}$ are linear, 
and the polynomials $R$ and $S$ do not depend on $\yv$,
successive applications of $\mathcal{R}$ and $\mathcal{S}$ to $J_3$ may be used to obtain 
the substitutions necessary to compute the integrals $\mathcal{I}_{a1c}$ and $\mathcal{I}_{a0c}$.
The procedure is detailed below in Sections \ref{subappend_IaUc}-\ref{subsec_Ia0c}.

\subsection{Substitutions for the integrals $\mathcal{I}_{a1c}$}
\label{subappend_IaUc}

The computation of the integrals $\mathcal{I}_{a1c}$ (with $a+c=2$) 
requires substitutions for the polynomials $H_{a0b}$ with $a+b=2$.
The application of $\mathcal{R}^2$, $\mathcal{S}^2$ 
and $\mathcal{R} \circ \mathcal{S}$  to $J_3$, respectively yield
\begin{align}
\label{eq_sub_ac_2_h200}
H_{200} + 2r H_{110} + r^2 H_{020} = &  P_{20} \, , \\
\label{eq_sub_ac_2_h002}
H_{002} + 2s H_{011} + s^2 H_{020} = &  P_{02} \, , \\
\label{eq_sub_ac_2_h101}
H_{101} + s H_{110} + r H_{011} + r s H_{020} = &  P_{11} \, ,
\end{align}
where the expressions of $P_{20}$, $P_{02}$, $P_{11}$ are given 
by Eqs. (\ref{eq_P20}-\ref{eq_P02}-\ref{eq_P11}).

\subsection{Substitutions for the integrals $\mathcal{I}_{a0c}$}
\label{subsec_Ia0c}

The computation of the integrals $\mathcal{I}_{a0c}$ (with $a+c=3$) 
requires substitutions for the polynomials $H_{a0b}$ with $a+b=3$.
The application of $\mathcal{R}^3$, $\mathcal{S}^3$, 
$\mathcal{R}^2 \circ \mathcal{S}$, $\mathcal{R} \circ \mathcal{S}^2$ to $J_3$,
respectively yield
\begin{align}
\label{eq_sub_ac_3_h300}
H_{300} + 3r H_{210} + 3r^2 H_{120} + r^3 H_{030} = &  P_{30} \, , \\
\label{eq_sub_ac_3_h003}
H_{003} + 3s H_{012} + 3s^2 H_{021} + s^3 H_{030} = &  P_{03} \, , \\
\label{eq_sub_ac_3_h201}
H_{201} + s H_{210} + 2r H_{111} + 2 r s H_{120} + r^2 H_{021} + r^2s H_{030} = &  P_{21} \, , \\
\label{eq_sub_ac_3_h102}
H_{102} + r H_{012} + 2s H_{111} + 2 r s H_{021} + s^2 H_{120} + r s^2 H_{030} = &  P_{12} \, ,
\end{align}
where the expressions of $P_{30}$, $P_{03}$, $P_{21}$, $P_{12}$ 
are given by Eqs. (\ref{eq_P30}-\ref{eq_P03}-\ref{eq_P21}-\ref{eq_P12}).

\section{Expressions of the function $G_0$ and $G_1$}
\label{sec_G0_G1}

After completing the square for $\tau$ in the exponential argument of $J_3$, 
the function $G_0$ (defined in Eq. \ref{eq_G0_general}) may be expressed in closed form as follows
\begin{align}
\begin{split}
\label{eq_G0_final}
G_0(\tilde{\lc},\zv) =  & \sqrt{\frac{\pi}{2}} \sqrt{\frac{\delta_3}{\delta_2}} \exp 
\left\{ - \frac{1}{2} \frac{\tilde{\lc}^2 - 2\rh \tilde{\lc} \zv + \zv^2 }{\delta_2} \right\} 
              \left[ 1 + \erf\left( \frac{\rhp}{\sqrt{2\delta_2\delta_3}} (-\rh \tilde{\lc} + \zv) \right) \right] \, ,
\end{split}
\end{align}
where $\delta_3$ has been defined in \eq{eq_d3}, and 
\begin{equation}
\delta_2 = 1 - \rh^2 \, 
\end{equation}
has been introduced to compact the expression. 
Then, an expression for the function $G_1$, defined in \eq{eq_G1_general}, 
may be readily obtained by substituting the prefactor $\tau$ (using Eq. \ref{eq_hernite_h010}) as follows:
\begin{equation}
\label{eq_G1_intermediate}
G_1(\tilde{\lc},\zv) = 
\int_0^{+\infty} \frac{1}{\delta_2} \left( \delta_3 H_{010}(\tilde{\lc},\yvi,\zeta) - \rh\rhp \tilde{\lc} + \rhp \zeta \right) J_3(\tilde{\lc},\yvi,\zv) \ {\rm d} \yvi \, ,
\end{equation}
leading to
\begin{equation}
\label{eq_G1_final}
G_1(\tilde{\lc},\zv) = 
\frac{\delta_3}{\delta_2} J_3(\tilde{\lc},0,\zeta) 
+ \frac{\rhp}{\delta_2} (-\rh \tilde{\lc} + \zeta) G_0(\tilde{\lc},\zeta) \, .
\end{equation}

\section*{Acknowledgements}

The author is grateful to Marc PREVOSTO for sharing 
his Monte Carlo code of second-order waves.

\bibliography{mybibfile}

\end{document}